\documentclass[aps,prd,nofootinbib]{revtex4}
\usepackage{epsfig,graphics,graphicx}

\usepackage{slashed}
\usepackage{amsmath,lscape,epsfig}
\usepackage{amsfonts}
\usepackage{amsmath}
\usepackage{amssymb}
\usepackage{graphicx}

\def\beq{\begin{equation}}
\def\eeq{\end{equation}}
\def\beqa{\begin{eqnarray}}
\def\eeqa{\end{eqnarray}}
\def\ban{\begin{eqnarray*}}
\def\ean{\end{eqnarray*}}
\def\bi{\begin{itemize}}
\def\ei{\end{itemize}}

\def\bpsi{\overline{\psi}}
\def\dslash{\slashed{\partial}}

\def\bpsi{\overline{\psi}}

\def\Epu{E_{{\bf p}_1}}
\def\Epd{E_{{\bf p}_2}}

\def\sumint{\hbox{$\sum$}\!\!\!\!\!\!\!\int}
\def\LMS{\Lambda_{\overline{\textrm{MS}}}}

\begin{document}

\title{Nonperturbative Yukawa theory at finite density and temperature}

\author{Eduardo S. {\sc Fraga}$^{a}$\footnote{fraga@if.ufrj.br}, 
Let\'\i cia F. {\sc Palhares}$^{a}$\footnote{leticia@if.ufrj.br}, and 
Marcus {\sc Benghi Pinto}$^{b}$\footnote{marcus@fsc.ufsc.br} }

\affiliation{$^{a}$Instituto de F\'{\i}sica, Universidade Federal do Rio de Janeiro \\ 
C.P. 68528, Rio de Janeiro, RJ, 21941-972, Brazil \\
$^{b}$Departamento de F\'{\i}sica, Universidade Federal de Santa
Catarina \\ C.P. 476, Florian\'opolis, SC,  88.040 - 900, Brazil}


\begin{abstract}

In-medium Yukawa theory is part of the thermodynamics of the Standard Model of particle physics 
and is one of the main building blocks of most effective field theories of 
fermionic systems. By computing its pressure we investigate the nonperturbative thermodynamics
at finite temperature and density using the optimized perturbation 
theory (OPT) framework. Our calculations are valid for arbitrary fermion and 
scalar masses, temperature, chemical potential, and not restricted to weak coupling.
The model is considered in the presence as well as in the absence of condensates.
Comparison with nonperturbative results  shows that second order perturbation 
theory (PT) fails in the first case but performs rather well when condensates 
are absent, even at high-temperature regimes.

\vspace{0.5cm}
PACS number(s): 11.10.Wx, 12.38.Cy, 11.15.Tk

\end{abstract}

\maketitle

\section{Introduction}

Matter under extreme conditions often requires the use of effective field theories in the 
description of its thermodynamical properties, independently of the energy scale under 
consideration. For instance, in the phenomenon of spontaneous symmetry breaking and 
temperature dependence of antiferromagnetic order in high $T_{c}$ superconductors described 
by a field-theoretic version of the Hubbard model \cite{hubbard} at relatively low energies, one has 
to incorporate the interaction of fermions with the crystal lattice via phonons, in a coarse 
graining of quantum electrodynamics. Moving up in energy scale, the phase structure of strong 
interactions and the quark-gluon plasma \cite{Rischke:2003mt}, investigated in high-energy 
heavy-ion collisions experiments \cite{QM} and in the observation of compact stars \cite{stars}, 
often demands simplifications of quantum chromodynamics, producing a variety of low-energy 
effective models \cite{Stephanov:2007fk}. 

Besides the major role it plays in the mechanisms of spontaneous symmetry breaking and mass 
generation in the Standard Model of particle physics, the Yukawa interaction stands out as one 
of the main ingredients in the construction of simplifying effective field theories to study the 
thermodynamics of systems under extreme conditions, especially if one imposes renormalizability. 
Although an effective theory does not require renormalizability to be consistent, a physically-motivated 
cutoff being usually more than satisfactory in this case, this feature can prove to be useful in the 
study of scale dependence and running via renormalization group methods \cite{Palhares:2008yq}. 

In this paper we investigate the full nonperturbative thermodynamics of the Yukawa theory at finite 
temperature and density by computing its pressure using the optimized perturbation theory (OPT) framework.
In our evaluations we consider contributions up to two loops, which include direct (Hartree like)
as well as exchange (Fock like) terms, so that the Yukawa thermodynamics can be investigated
in the presence of condensates and also in their absence. In the first case, besides  
OPT and ordinary perturbation theory (PT) we shall also perform a mean field (MFT) evaluation.
This is an important step in establishing the OPT reliability since the reader will see 
that when exchange contributions are neglected OPT exactly reproduces MFT results 
which can be considered ``exact'' in this (large $N$) limit \footnote{
The reliability of the OPT framework applied to symmetric and broken phases was studied
previously in the context of $\phi^4$ scalar theories (see for instance Ref. \cite{Farias:2008fs}).
}.
We present results  which are 
valid for arbitrary fermion and scalar masses, temperature, chemical potential, and coupling. 
The region of large values of the coupling is particularly interesting and useful, since several effective 
field theory models in particle and nuclear physics exhibit Yukawa coupling constants that are much 
larger than one, as is the case, for instance, in the linear sigma 
model \cite{GellMann:1960np}, frequently used in the description of the chiral phase 
transition \cite{Bilic:1997sh,quarks-chiral,ove,Scavenius:1999zc,Caldas:2000ic,Scavenius:2000qd,Scavenius:2001bb}, 
and in pion-nucleon models extracted from chiral Lagrangians \cite{Weinberg:1978kz,vanKolck:1999mw}.

When condensates are present we show that, as expected, PT has a poor performance since it cannot 
resum direct (tadpole) contributions associated with symmetry breaking. On the other hand, OPT 
improves over MFT by also incorporating exchange terms in a nonperturbative fashion.

In the absence of condensates and in the regime of very small coupling our numerical findings can be verified in 
the limit of vanishing temperature by comparison to exact analytic two-loop perturbative results 
previously obtained by some of us for the equation of state of cold and dense Yukawa theory within 
the $\overline{\rm MS}$ scheme \cite{Palhares:2008yq}. In Ref. \cite{Palhares:2008yq}, the 
two-loop momentum integrals were computed analytically for {\it arbitrary} fermion and scalar masses, 
the final result being expressed in terms of well-known special functions, which provides us with 
a solid and  clear reference in this limit. 

Furthermore, this comparison also provides another way of testing the idea 
that perturbation theory at high density and zero temperature in the symmetric phase 
is much better behaved than its 
converse \cite{Palhares:2008yq,Fraga:2001id,tony,andersen,Braaten:2002wi,Palhares:2007zz}, which 
has well-known severe infrared problems \cite{kapusta-gale}. Our analysis shows that, surprisingly, 
second-order PT agrees quite well with the OPT nonperturbative results up to values of the 
coupling of order one in the case of cold and dense as well as hot and dense Yukawa theory.

The framework of OPT \cite{OPT}, also known as the linear delta expansion, 
is an example of a variational method that implements the resummation of certain classes of 
Feynman diagrams, incorporating nonperturbative effects in the computation of the thermodynamic 
potential (for related methods, see Refs. \cite{OPT2}). 
It has been successfully applied to the study of many different physical situations, such as mapping 
the phase diagram of the 2+1 dimensional Gross-Neveu model \cite{Kneur:2007vj}, where a previously undetermined 
``liquid-gas'' phase has been located,  and determining the critical 
temperature for Bose-Einstein condensation in dilute interacting atomic gases \cite{BEC}.
Some early applications at finite temperature can be found in Refs. \cite{OPT3}.

The paper is organized as follows. In Section II we present the in-medium Yukawa theory and set 
up the notation of the paper. Section III contains a sketch of the derivation of the perturbative 
thermodynamic potential at finite temperature and chemical potential. In Section IV we apply the 
OPT machinery to the evaluation of the nonperturbative thermodynamic potential in the Yukawa 
theory. The thermodynamics of the Yukawa theory, in the presence of condensates, is
considered in Section V where the results from PT, OPT and MFT are contrasted. 
In Section VI we consider the symmetric case, where condensates are absent. 
Comparing the results from OPT and PT, we show that the latter performs well also for 
extreme temperature values up to two-loop order.
Section VII contains our conclusions. Technical details involved in the calculation of
the vacuum contributions and direct terms in the two-loop thermodynamic potential are left for appendices
\footnote{The technicalities concerning Matsubara sums, renormalization, and the analytic 
evaluation of in-medium momentum integrals in the calculation of the medium contributions up to two loops
were extensively addressed in the appendix of Ref. \cite{Palhares:2008yq}.}.


\section{In-medium Yukawa theory}

In what follows, we consider a gas of $N_{F}$ flavors of massive spin-$1/2$ fermions 
whose interaction is mediated by a massive real scalar field, $\phi$, with an interaction 
term of the Yukawa type, so that the Lagrangian has the following general form:
\begin{equation}
\mathcal{L}_{Y}
= \mathcal{L}_{\psi}+
\mathcal{L}_{\phi}
+\mathcal{L}_{int} \, ,
\label{Lyukawa}
\end{equation}
where
\begin{eqnarray}
\mathcal{L}_{\psi} &=& \sum_{\alpha=1}^{N_F} \bpsi_{\alpha}\left( i\dslash -m \right)
\psi_{\alpha} \, ,
\\
\mathcal{L}_{\phi} &=& \frac{1}{2} (\partial_{\mu}\phi)(\partial^{\mu}\phi)-
\frac{1}{2} m_{\phi}^2 \phi^2
-\lambda_3\phi^3-\lambda\phi^4 \, ,
\\
\mathcal{L}_{int} &=& \sum_{\alpha=1}^{N_F} g~\bpsi_{\alpha}\psi_{\alpha}\phi \, . 
\label{Lint}
\end{eqnarray}
Here, $m$ and $m_{\phi}$ are the fermion and boson masses, respectively, assuming all the 
fermions have the same mass, for simplicity. The Yukawa coupling is represented by $g$; 
$\lambda_3$ and $\lambda$ are bosonic self-couplings allowed by renormalizability. 
Here we choose $\lambda_3=\lambda=0$  disregarding  bosonic self-interactions which will be treated in a
future work \cite{future}.

We work in the imaginary-time Matsubara formalism of finite-temperature 
field theory, where the time dimension is compactified and associated with the inverse 
temperature $\beta=1/T$ \cite{kapusta-gale}. In this approach, one has to impose 
periodicity (anti-periodicity) for the bosonic (fermionic) fields in the imaginary time $\tau$, 
in order to satisfy the spin-statistics theorem. Therefore, only specific discrete Fourier modes 
are allowed, and integrals over the zeroth component of four-momentum are replaced by 
discrete sums over the Matsubara frequencies, denoted by $\omega^B_n=2n\pi T$ for bosons 
and $\omega^F_n=(2n+1)\pi T$ for fermions, with $n$ integer. Nonzero density effects are included 
by incorporating the constraint of conservation of the fermion number via a shift in the zeroth component 
of the fermionic four-momentum $p^0=i\omega_n^F\mapsto p^0=i\omega_n^F+\mu$, $\mu$ being 
the chemical potential.

\vspace{0.4cm}
\begin{figure}[htb]
\includegraphics[width=17cm]{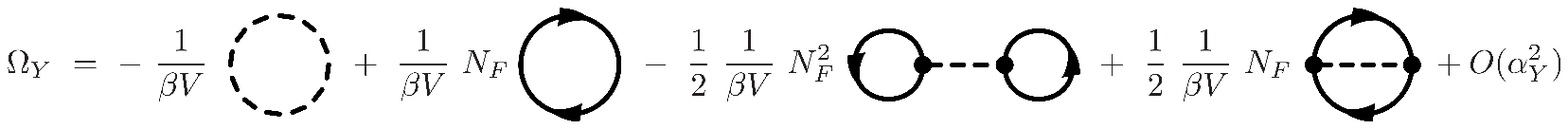}
\caption{Perturbative thermodynamic potential for the Yukawa theory to two loops. 
Solid lines represent fermions and dashed lines stand for  bosons. 
Here we omit the diagrams containing counterterms. The third (tadpole) contribution can
be neglected when we consider the symmetric case.}
\label{OmegaY-fig}
\end{figure}

From the partition function written in terms of the euclidean action for the lagrangian (\ref{Lyukawa}), 
 $Z_{Y}(T,\mu) = {\rm Tr} ~\exp(-S_Y)$, one derives the perturbative series for the thermodynamic 
 potential $\Omega_{Y}\equiv - (1/\beta V) \ln Z_Y$ \cite{kapusta-gale}:
\begin{equation}
\Omega_{Y}=
-\frac{1}{\beta V} \ln Z_0-
\frac{1}{\beta V}\ln \left[ 1+\sum_{\ell=1}^{\infty} \frac{(-1)^\ell}{\ell !}
\langle S_{int}^{\ell} \rangle_0 \right] \, ,
\end{equation}
where $V$ is the volume of the system, $Z_0$ is the partition function 
of the free theory and $S_{int}$ represents the euclidean interaction action. 
Notice that Wick's theorem implies that only even powers in the above 
expansion survive, yielding a power series in $\alpha_Y\equiv g^2/4\pi$. 
However, at finite temperature the perturbative expansion also contains 
odd powers of $g$ coming from resummed contributions of the zeroth 
Matsubara mode for bosons, such as in the case of the plasmon 
contribution \cite{kapusta-gale}. In the zero-temperature limit this is not 
the case and, even including hard-dense-loop corrections, only $g^{2\ell}$ terms 
are modified \cite{andersen}. Since we restrict our analysis to two 
loops (the resummed OPT calculation also departs from a two-loop perturbative 
setting), we are not concerned with this issue. Omitting the diagrams representing 
counterterms, the perturbative thermodynamic potential is shown, diagrammatically, in 
Fig. \ref{OmegaY-fig}. 
The first two diagrams correspond to the free gas, 
while the last two represent interaction terms: the third diagram is a contribution of
the direct type and the fourth is of the exchange type. Notice that the third diagram,
which contains tadpoles, belongs to the one-particle-reducible class, that
does not contribute in the absence of condensates. It will therefore
be neglected in Section VI, where only the symmetric case is considered.

\section{Perturbative Thermodynamic Potential at finite $\mu$ and $T$}

Considering the thermodynamic potential only up to the first non trivial (two-loop) contribution, 
which corresponds to direct plus exchange terms, one must evaluate the following $O(g^2)$ quantity:
\begin{equation}
\Omega_Y =\frac{i}{2}  ~\sumint_{K}
\ln \left[K^2 - m^2_ \phi \right]
+i  ~\sumint_{P} {\rm Tr}\ln \left[\not \! P - m \right] 
- \frac{i}{2}  ~\sumint_{P}  {\rm Tr} \left[\frac {\Sigma_{\rm dir}(0)+\Sigma_{\rm exc}(P)}{(\not \! P - m)}\right] \,\,,
\label{PT}
\end{equation}
where the trace is to be performed over the Dirac structure, and the $4$-momenta are given 
in terms of the Matsubara frequencies and the $3$-momenta for fermions, 
$P=\left( p^0=i\omega_{n}^{F}+\mu \, ,\,  {\bf p}\right)$, and bosons, 
$K=\left( k^0=i\omega_{\ell}^{B}\, ,\, {\bf k} \right)$. We choose the metric tensor signature 
$g^{\mu\nu}={\rm diag}(+,-,-,-)$, and the following short-hand notation for sum-integrals:
\begin{equation}
\sumint_{P} = T \sum_{n} \int \frac{d^{3}{\bf p}}{(2\pi)^{3}} \, .
\end{equation}
The direct contribution is given by
\begin{equation}
\Sigma_{\rm dir}(0)= -i \left (\frac{g}{m_\phi} \right)^2 ~\sumint_{Q}
{\rm Tr}\frac{1}{ [\not \! Q - m]} \, ,
\end{equation}
and, the exchange self-energy is given by the following sum-integral over fermionic momenta:
\begin{equation}
\Sigma_{\rm exc}(P)= i g^2 ~\sumint_{Q}
\frac{1}{ [\not \! Q - m][(P-Q)^2 - m^2_\phi]} \, .
\end{equation}

Computing the traces at $\mu \ne 0$ and $T \ne 0$ (see Ref. \cite {Palhares:2008yq}
for details concerning the exchange term), one can write the 
thermodynamic potential as $\Omega_Y=\Omega_Y^{\rm vac}+\Omega_Y^{\rm med}(T,\mu)$, 
where the vacuum contribution prior to renormalization has the form
\begin{eqnarray}
\Omega_Y^{\rm vac}&=& \frac{1}{2} \int \frac {d^3 {\bf k}}{(2\pi)^3} 
\omega_{\bf k} - 2N_{\rm F} \int \frac {d^3 {\bf p}}{(2\pi)^3} E_{\bf p} 
-\frac{g^2N_F^2}{2m_{\phi}^2} \left[2m\int \frac{d^{3}{\bf p}}{(2\pi)^{3}}  \frac{1}{E_{\bf p}}\right ]^2
- g^2 \frac {N_{\rm F}}{4} 
\int \frac {d^3 {\bf p}_1}{(2\pi)^3} \frac {d^3 {\bf p}_2}{(2\pi)^3} 
\frac { {\overline{\cal J}_-}(E_+-\omega_{12})}{\omega_{12}E_{{\bf p}_1}E_{{\bf p}_2}} 
\nonumber\\
&=&
-B(m_{\phi})+4N_F B(m)+\Omega^{{\rm dir}}_{{\rm vac}}(m,m_{\phi})+\Omega^{{\rm exc}}_{{\rm vac}}(m,m_{\phi})
\, , \label{OmegaY(0,0)}
\end{eqnarray}
\noindent where $\omega_{\bf k}=({\bf k}^{2}+m_{\phi}^{2})^{1/2}$,  
$E_{\bf p}=({\bf p}^{2}+m^{2})^{1/2}$, 
$E_{\pm} \equiv \Epu \pm \Epd$, 
$\omega_{12} \equiv ( |{\bf p}_1-{\bf p}_2|^2+m_{\phi}^2)^{1/2}$, 
and we have defined
\begin{equation}
\overline{\mathcal{J}}_{\pm} \equiv 
-2~\frac{m^2-{\bf p}_1\cdot {\bf p}_2\pm \Epu\Epd}{E_{\mp}^2-\omega_{12}^2}
= 1-\frac{4m^2-m_{\phi}^2}{E_{\mp}^2-\omega_{12}^2} \, .
\label{barJpm}
\end{equation}
Notice that Eq. (\ref{OmegaY(0,0)}) contains zero-point energy divergent terms 
which are $T$- and $\mu$- independent only within the PT approach in which case they can be conveniently absorbed by the usual vacuum subtraction which normalizes  the pressure so that it vanishes at $\mu=0$.
In practice this means that, as done in Ref. \cite {Palhares:2008yq}, one does not have to care to their explicit evaluation and renormalization. However, within the other approaches (OPT and MFT)  considered here  the PT bare mass, $m$, is replaced by effective ($T$- and $\mu$- dependent) masses within the same diagrams. Therefore, we must renormalize those contributions
appropriately. Within the $\overline {\rm MS}$ subtraction scheme, the fully renormalized vacuum term can be written 
as  (see Appendices \ref{ApRen} and \ref{Apdir} for details)
\begin{equation}
\Omega_Y^{\rm vac}=-B^{{\rm REN}}(m_{\phi})+4N_F B^{{\rm REN}}(m)+\Omega^{{\rm dir,REN}}_{{\rm vac}}(m,m_{\phi})+
\Omega^{{\rm exc,REN}}_{{\rm vac}}(m,m_{\phi}) \, ,
\end{equation}
where we have defined the following functions of the masses:
\begin{eqnarray}
B^{{\rm REN}}(M)&\equiv& \frac{M^4}{64\pi^2}\left[ \frac{3}{2}+\log\left( \frac{\Lambda^2}{M^2} \right) \right] \, ,
\\
\Omega^{{\rm dir,REN}}_{{\rm vac}}(m,m_{\phi})&\equiv&-\frac{g^2 N_F^2}{2 m_\phi^2} \left \{ \frac{m^3}{(2 \pi)^2} \left [  1+ \ln \left(\frac{\Lambda^2}{m^2} \right ) \right ] \right \}^2
\\
\Omega_{{\rm vac}}^{{\rm exc,REN}}(m,m_{\phi}) &\equiv&  
N_F \frac{g^2}{2} \frac{m^4}{64\pi^{4}}~
\Bigg\{ 
v_1\left( \frac{m_{\phi}^2}{4m^2} \right)
+\left[ \gamma +\log\left( \frac{\Lambda^2}{m^2}\right)
 \right]~v_2\left( \frac{m_{\phi}^2}{4m^2} \right)
+
\frac{1}{2}~\left[ \gamma +\log\left( \frac{\Lambda^2}{m^2}\right)\right]^2
~v_3\left( \frac{m_{\phi}^2}{4m^2} \right)
+
\nonumber\\
&&\quad +
6~\alpha\left(m^2\right)-\frac{m_{\phi}^4}{m^4}\left( 1-6\frac{m^2}{m_{\phi}^2} \right)
\alpha\left(m_{\phi}^2\right)
\Bigg\} \, .
\end{eqnarray} 
Here $\gamma$ is the Euler constant, $\Lambda$ is the renormalization scale in the $\overline {\rm MS}$ 
scheme, and the functions $v_i(z)$ and $\alpha\left(m^2\right)$ are defined in Appendices \ref{ApRen} and \ref{Apdir}.

Adding the direct term to the free gas plus exchange medium-dependent terms evaluated in Ref. \cite {Palhares:2008yq} one 
obtains
\begin{eqnarray}
\Omega_Y^{\rm med}(T,\mu)
&=& \frac{T^4}{2\pi^2} \int_0^\infty z^2dz \log[1-e^{-{\omega_z}}] 
\nonumber \\
&-& T^4 \frac{N_{\rm F}}{\pi^2}\int_0^\infty z^2dz\left \{  \log[1+e^{-(E_z -\mu/T)}] +\log[1+e^{-(E_z +\mu/T)}] \right \}
\nonumber \\
&-& g^2 T^2 \frac{N_{\rm F}^2 m^4}{(4 \pi^4) m_\phi^2}\left [  1+ \ln \left(\frac{\Lambda^2}{m^2} \right ) \right ]\left \{\int_0^{\infty} z^2 dz \frac{1}{[z^2 + m^2/T^2]^{1/2}} \left [
\frac{1}{[1+e^{(E_{z} -\mu/T)}]} +\frac{1}{[1+e^{(E_{z} +\mu/T)}]} \right]\right \} 
 \nonumber \\
&-& g^2 T^4 \frac{ N_{\rm F}^2 m^2}{(2 \pi^4) m_\phi^2}\left \{\int_0^{\infty} z^2 dz \frac{1}{[z^2 + m^2/T^2]^{1/2}} \left [
\frac{1}{[1+e^{(E_{z} -\mu/T)}]} +\frac{1}{[1+e^{(E_{z} +\mu/T)}]} \right]\right \}^2 
\nonumber \\
&-& g^2 T^2 m^2 \frac{N_{\rm F}}{(2\pi)^4} ~\alpha_1 \int_0^\infty z^2dz \left [ \frac{N_f(1)}{E_z}
 \right ] \nonumber \\
&-& g^2 T^2 \frac{N_{\rm F}}{(2\pi)^4}~( \alpha_2 + 3\alpha_3)
\int_0^\infty z^2dz \left [ \frac{n_b(\omega_z)}{\omega_z} \right ]
\nonumber \\
&+& g^2 T^4 \frac{N_{\rm F}}{2(2\pi)^4}\int_0^\infty z^2dz y^2 dy\int_{-1}^1du_{zy} 
\frac{1}{E_z E_y}\left [ {\tilde {\cal J}_+}\Sigma_1 +{\tilde {\cal J}_-}\Sigma_2 \right ]
\nonumber \\
&+& g^2 T^4 \frac{N_{\rm F}}{(2\pi)^4}\int_0^\infty z^2dz x^2 dx \int_{-1}^1du_{zx}
\frac{1}{\omega_{x} E_z E_{zx}} \left [{\tilde {\cal K}_-}{\tilde {E}_+} - 
{\tilde {\cal K}_+}{\tilde {E}_-} \right ] n_b(\omega_{x}) N_f(1) 
\,\,,
\label{Omega(T,mu)}
\end{eqnarray}
%
where, in order to perform numerical investigations, we have defined the following dimensionless quantities: 
$\omega_z^2= z^2 + {m_{\phi}}^2/T^2$, $E_z^2= z^2+m^2/T^2$, ${\tilde {E}_\pm}=E_z \pm E_{zx}$, 
$E_{zx}^2= x^2+z^2+2xz ~u_{zx} + m^2/T^2$, and
\begin{eqnarray}
{\tilde {\cal J}_\pm}&=& 1 + \frac { 4(m/T)^2 - (m_\phi/T)^2} { (E_z\mp E_y)^2 - {\omega_{zy}}^2}\,\,,
\\
{\tilde {\cal K}_\pm}&=& 1 + \frac { 4(m/T)^2 - (m_\phi/T)^2} { {\tilde{E}_{\mp}}^2 - {\omega_{x}}^2}\,\,,
\\
N_f(1) &=& n_f(E_z+\mu/T ) +n_f(E_z-\mu/T ) \, ,
\label{Nfs}
\\  
\Sigma_1 &=& n_f(E_z+\mu/T )~n_f(E_y+\mu/T )+n_f(E_z-\mu/T )~n_f(E_y-\mu/T ) \, ,
\label{Sigma1}
\\
\Sigma_2 &=& n_f(E_z+\mu/T )~n_f(E_y-\mu/T )+n_f(E_z-\mu/T )~n_f(E_y+\mu/T ) \, ,
\label{Sigma2}
\\
\alpha_1&=&
-4 \frac{m_{\phi}}{m}\left( 1-\frac{m_{\phi}^2}{4m^2} \right)^{\frac{3}{2}}
\left\{ \tan^{-1}\left[ \sqrt{\frac{1}{\frac{4m^2}{m_{\phi}^2}-1}} \right]
+\tan^{-1}\left[ \frac{\frac{1}{2}-\frac{m_{\phi}^2}{4m^2}}{\sqrt{ \frac{m_{\phi}^2}{4m^2} }\sqrt{1-\frac{
m_{\phi}^2}{4m^2}}} \right] \right\}
\nonumber \\
&& +\frac{7}{2}-\frac{m_{\phi}^2}{2m^2}-\frac{3}{2}\log\left( 
\frac{m^2}{\Lambda^2} \right)+\frac{m_{\phi}^2}{m^2}\left( \frac{3}{2}
-\frac{m_{\phi}^2}{4m^2} \right)
\log\left( \frac{m^2}{m_{\phi}^2} \right)
\, ,
\label{alpha1res}
\\
\alpha_2&=& m^2- \frac{1}{6}m_{\phi}^2
\, ,
\\
\alpha_3&=& \frac{2}{3}\left[ 2m^2-\frac{5}{12}m_{\phi}^2 \right]-
\frac{1}{3}m_{\phi}^2\left( \frac{4m^2}{m_{\phi}^2}-1 \right)^{\frac{3}{2}}
\tan^{-1}\left[ \frac{1}{\sqrt{\frac{4m^2}{m_{\phi}^2}-1}} \right]
-\left( m^2-\frac{m_{\phi}^2}{6} \right)\log\left( \frac{m^2}{\Lambda^2} \right)
\, ,\label{alpha3res}
\end{eqnarray}
with $\omega_{zy}^2=z^2+y^2+2zy ~u_{zy}+{m_{\phi}}^2/T^2$, $x=k/T$, $z=p_1/T$, $y=p_2/T$, $n_b(x)=[\exp(x)-1]^{-1}$, and 
$n_f(x)=[\exp(x)+1]^{-1}$.
%

%
Given the perturbative expressions obtained in this section, we can compute the thermodynamic 
potential $\Omega_{Y}$ in the OPT framework, which corresponds to resumming all dressed 
diagrams of the direct and exchange types, generating in practice an effective mass for the fermions.

\section{Nonperturbative thermodynamic potential in the OPT framework}

Following the standard procedure, the OPT framework \cite{OPT} can be implemented 
in the Yukawa theory as
\begin{equation}
\mathcal{L}_{\rm OPT} =  \bpsi_{\alpha}\left[ i\dslash -(m+ \eta^*) \right]
\psi_{\alpha} + \delta \left[\frac{1}{2} (\partial_{\mu}\phi)^2-
\frac{1}{2} m_{\phi}^2 \phi^2\right]+ \delta ~g\phi \bpsi_{\alpha}\psi_{\alpha} \, ,
\label{Lint}
\end{equation}
where a sum over flavors is implied, and $\eta^* = \eta(1-\delta)$. As one can easily notice from the 
deformed Lagrangian above, at $\delta=0$ the theory is an exactly solvable theory of massive fermions 
even in the case when, originally, $m=0$. In this particular case, the {\it arbitrary}  mass parameter, 
$\eta$, works as an infrared regulator which proves to be very useful in studies related to chiral symmetry 
breaking. We should remark that, due to our choice, the meson sector disappears at $\delta=0$ since 
there is no need for their mediation in this case. Finally, when $\delta=1$ the {\it original}, interacting 
theory is recovered, so that our $\mathcal{L}_{\rm OPT}$ interpolates between a free (exactly solvable) 
fermionic theory and the original one.

The Feynman rules generated by the interpolating theory are trivially obtained from the original ones: 
$m \mapsto m+\eta^*$, $g \mapsto \delta g$, and the bosonic propagator receives a $1/\delta$ factor. 
It is important to notice that the interpolation does not change the polynomial structure of the 
original theory and hence does not spoil renormalization, as proved in Ref. \cite {PRDMR}. Now, a 
physical quantity such as $\Omega_Y^{\rm OPT}$ is {\it perturbatively} evaluated in powers of the 
dummy parameter, $\delta$, which is formally treated as small during the intermediate steps of the 
evaluation. At the end one sets $\delta=1$, in a procedure which is analogous to the one used within 
the large-$N$ approximation, where $N$ is formally treated as a large number and finally set to its 
finite value at the end. The result, however, depends on the arbitrary parameter, $\eta$, which can be 
fixed by requiring that $\Omega_Y^{\rm OPT}$ be evaluated at the point where it is less sensitive to 
this parameter \cite{PMS} (Principle of Minimal Sensitivity, PMS). This can be accomplished by requiring
\begin{equation}
\frac{d \Omega_Y^{\rm OPT}}{d \eta}{\Big |}_{\eta={\bar \eta}, \delta=1}=0 \,\,.
\label{pms}
\end{equation}
In general, the optimum value $\bar\eta$ becomes a function of the original couplings via self-consistent 
equations, generating nonperturbative results. 

Now we can apply the method to the case of the two-loop perturbative $\Omega_Y$ given in 
Eq. (\ref {PT}). Using the OPT replacements for the Feynman rules and expanding $\eta^*$ to 
order $\delta$, one obtains the first nontrivial result
\begin{eqnarray}
\Omega_Y^{\rm OPT} &=& \delta ~\frac{i}{2}  \sumint_{K} 
\ln \left[ K^2 - m^2_ \phi \right]+i  \sumint_{P} 
{\rm Tr}\ln \left[\not \! P - (m+\eta) \right] 
+\delta~ i  \sumint_{P} 
{\rm Tr} \left[\frac {\eta}{[\not \! P - (m+\eta)]}\right] 
\nonumber \\
&& 
-\delta ~\frac{i}{2}  \sumint_{P} 
{\rm Tr} \left[\frac {\Sigma_{\rm dir}(0,\eta)+\Sigma_{\rm exc}(P,\eta)}{[\not \! P - (m+\eta)]}\right] 
\,, \nonumber \\
\label{TPopt}
\end{eqnarray}
where
\begin{equation}
\Sigma_{\rm dir}(0,\eta)= -i \left (\frac{g}{m_\phi} \right)^2 ~\sumint_{Q}
{\rm Tr}\frac{1}{ [\not \! Q - (m+\eta)]} \, 
\end{equation}
and
\begin{equation}
\Sigma_{\rm exc}(P,\eta)= i g^2 \sumint_{Q}
\frac{1}{ [\not \! Q - (m+\eta)][(P-Q)^2 - m^2_\phi]} \, .
\end{equation}
Notice that, when compared to the second-order perturbative result, Eq. (\ref {TPopt}) displays an extra contribution 
given by the third term on its right-hand side which represents a one-loop graph with a $\delta \eta$ insertion.
Now, using Eq. (\ref {pms}) one arrives at
\begin{eqnarray}
0&=&-i  \sumint_{P} {\rm Tr} \left[\frac {1}{[\not \! P - (m+{\bar \eta})]}\right]
+\delta~ i  \sumint_{P} {\rm Tr} \left[\frac {1}{[\not \! P - (m+{\bar \eta})]}\right]
+\delta~ i  \sumint_{P} {\rm Tr} \left[\frac {\bar \eta}{[\not \! P - (m+\eta)]^2}\right]
\nonumber \\
&& 
-\delta~ i   \sumint_{P} {\rm Tr} 
\left[\frac {\Sigma_{\rm dir}(0,{\bar \eta})+\Sigma_{\rm exc}(P,{\bar \eta})}{[\not \! P - (m+{\bar \eta})]^2}\right] \,, \nonumber \\
\end{eqnarray}
where, to obtain the last term, we have used a redefinition of momenta $p \mapsto q, 
q \mapsto p$ at an intermediate step. Setting $\delta=1$, one has a nontrivial, 
coupling-dependent self-consistent integral relation for the optimum mass parameter 
involving the self-energy given by
\begin{equation}
 i   \sumint_{P} {\rm Tr} 
\left[\frac {{\bar \eta}-\Sigma_{\rm dir}(0)-\Sigma_{\rm exc}(P,{\bar \eta})}{[\not \! P - (m+{\bar \eta})]^2}\right]
=0 \, .
\label{etapms}
\end{equation}
Since the OPT propagator has an infinite number of $\eta$ 
insertions, one can see that the optimization will resum exchange graphs in a nonperturbative way. 
It is very interesting to notice that when exchange terms are neglected 
${\bar \eta} = \Sigma_{\rm dir}(0)=-g\langle \phi \rangle_0 $, where $\langle \phi \rangle_0$
represents the scalar condensate, satisfying the MFT self-consistent relation for the 
effective mass (obviously, when $g\to 0$ the OPT results agree with the free gas case).
This type of result is consistent with applications of OPT to different
types of theories \cite {optmft} and  illustrates the way OPT works.
Notice that within the Hartree approximation one also adds and subtracts a mass 
term which is determined self-consistently. The basic difference is that the topology 
of this term is fixed from the start: direct terms in the Hartree approximation,
direct plus exchange in the Hartree-Fock approximation. Within OPT the effective
mass $m+\eta$ is {\it arbitrary} from the start, its optimum form being determined
by the topology of the contributions considered in the perturbative evaluation of a
physical quantity such as the thermodynamic potential considered here.

In the next two sections we shall study the pressure numerically, using different 
approximations in several different situations. With this aim we set 
$m=0.1 \LMS$, $m_\phi=m/2$ in our numerical routines \footnote {The value $m_\phi=0$,
which is of particular interest in situations motivated by QCD, will also be considered
in subsection \ref{mphi=0}.}. The temperature and chemical potential ranges considered
cover $0 \to 10m$ while the coupling values cover $0 \to \pi$.

\section{Results in the presence of a scalar condensate }

\begin{figure}[htb]

\includegraphics[width=7cm]{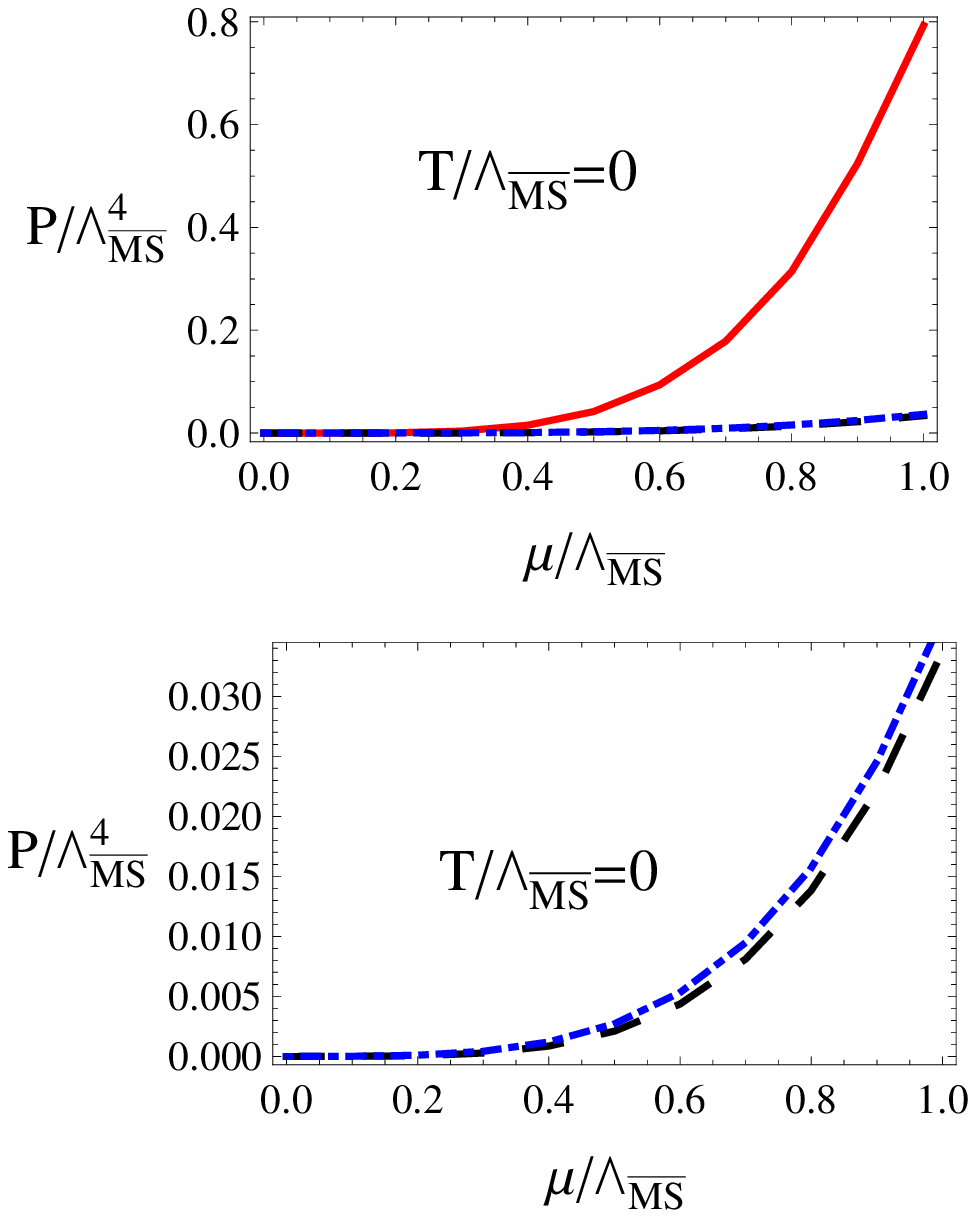}
\caption{
The pressure, $P/\LMS^4$,  as a function of $\mu/ \LMS$ at $T=0$. 
Top: results from PT (continuous line), MFT (dashed line) and OPT (dot-dashed line). 
Bottom: differences, due to exchange terms, between OPT and MFT. }
\label{DiExTzero}

\end{figure}

\begin{figure}[htb]
\includegraphics[width=7cm]{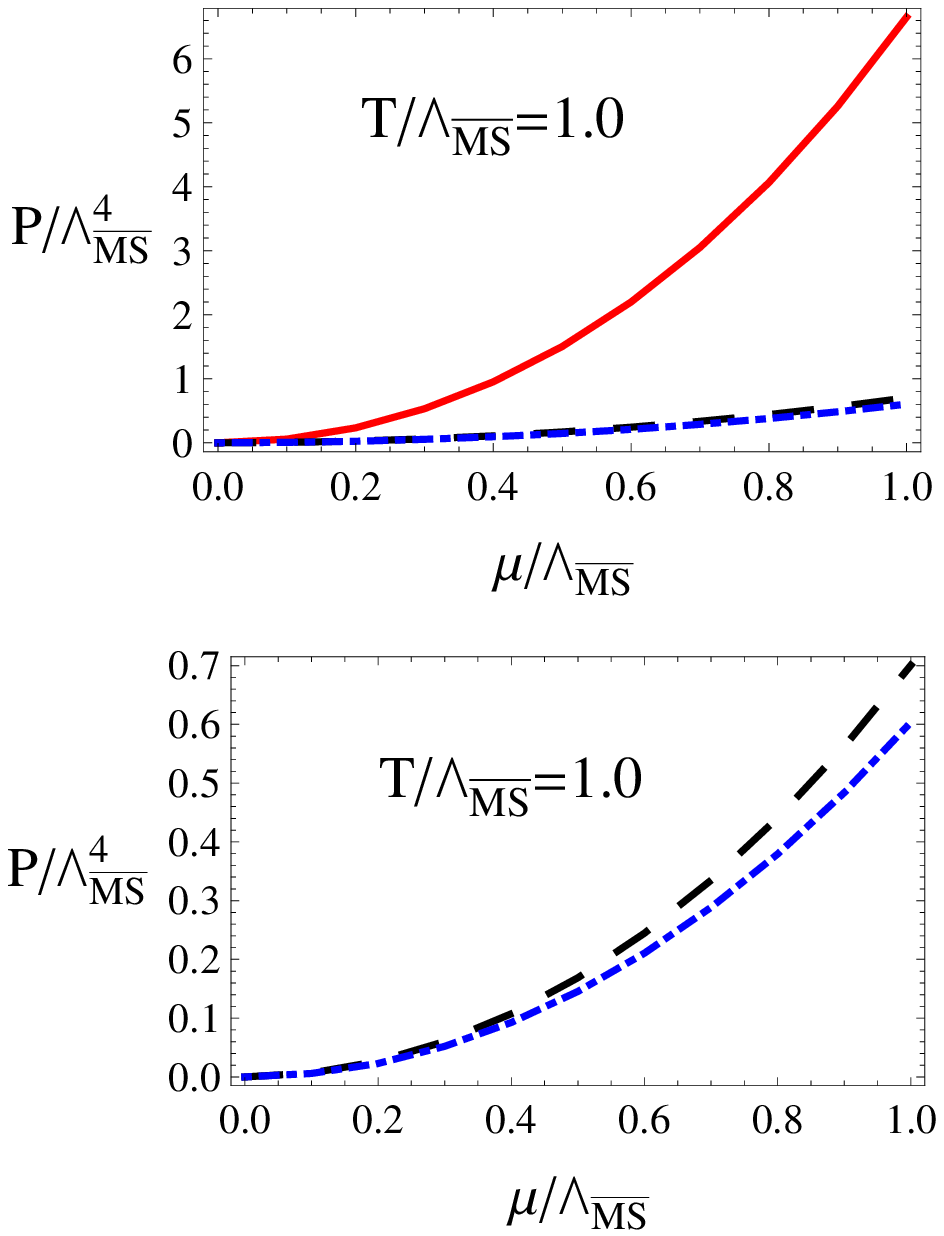}
\caption{The pressure, $P/\LMS^4$,  as a function of $\mu/ \LMS$ at $T=1.0 \LMS$. Top: results from PT (continuous line), MFT (dashed line) and OPT (dot-dashed line).
Bottom: differences, due to exchange terms, between the OPT and MFT. }
\label{DiExTum}
\end{figure}

In this section we compare the results generated by PT, OPT and MFT when the scalar
condensate represented by the direct (one-particle reducible) terms are considered. 
This case is also interesting because one can use the well-established MFT to 
analyze the results provided by OPT and PT. As already mentioned, when exchange
contributions are not considered in OPT, MFT results are {\it exactly}
reproduced since in this case both theories employ the same effective mass 
$M_{\rm eff}= m+\Sigma^{\rm MFT}_{\rm dir}(0)= m+ {\bar \eta}$. Each approximation
considers different two-loop contributions: PT takes all graphs shown in Fig. \ref {OmegaY-fig}
into account, while  OPT considers all plus the extra one-loop fermionic graph with the 
$\delta \eta$ insertion. In practice, as shown in the previous section, the MFT 
result is quickly recovered from the OPT one by neglecting the exhange term.
Figure \ref{DiExTzero} shows the pressure as a function of $\mu$ at $T=0$.
As one can see, PT predicts very high values for the pressure as $\mu$ increases, in disagreement
with MFT and OPT results. Since OPT agrees exactly with MFT when exchange
terms are neglected, one can also see in this figure the effects of resumming
exchange contributions: they yield slightly higher values of the pressure for increasing $\mu$.
Figure  \ref {DiExTum} shows the same situation but
at a high temperature. In this case, the OPT-predicted pressure values are
smaller than the MFT ones as $\mu$ increases.

\begin{figure}[htb]
\includegraphics[width=7cm]{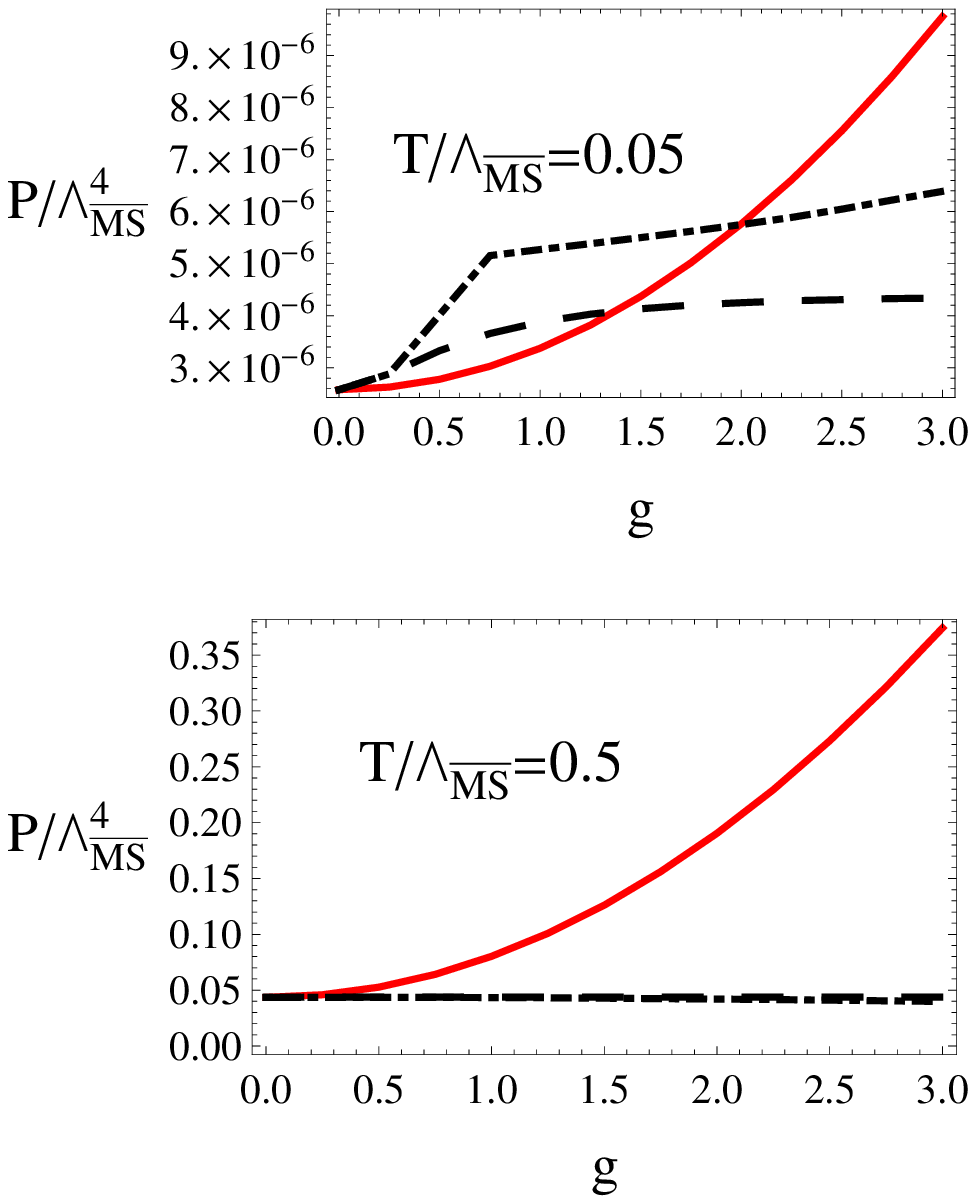}
\caption{The pressure, $P/\LMS^4$,  as a function of $g$ for $\mu=T=0.05 \LMS$ (top figure)  and $\mu=T=0.5 \LMS$ (bottom figure). PT corresponds to the continuous line, MFT to the dashed line and OPT to the dot-dashed line.}
\label{PvsgDEx}
\end{figure}

\begin{figure}[htb]
\includegraphics[width=7cm]{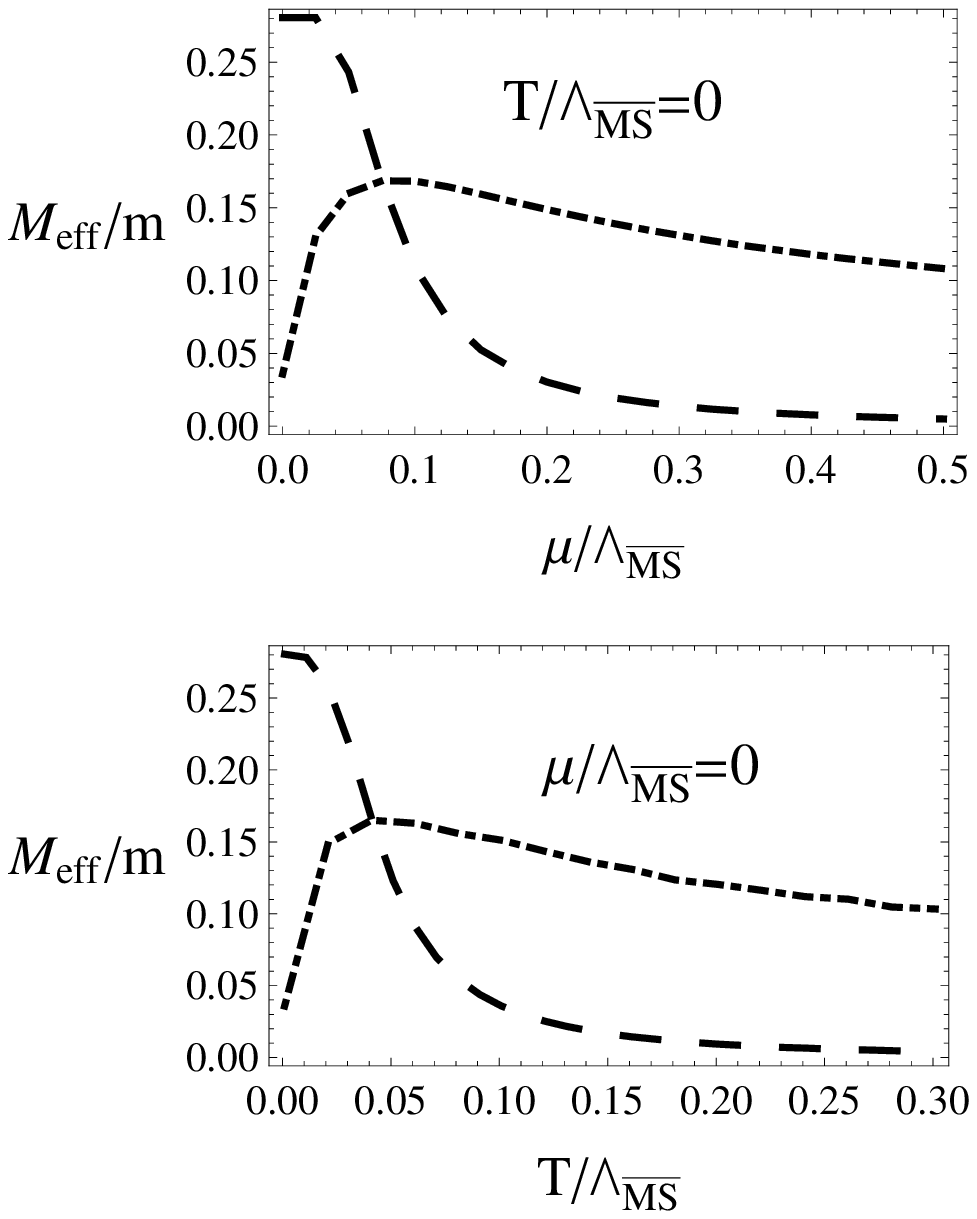}
\caption{The OPT (dot-dashed line) and MFT (dashed line) effective masses ($m + {\bar \eta}$ and $m+ \Sigma_{\rm dir}(0)$) in units of $m$ for $g=\pi$. 
Top: $M_{\rm eff}/m$ as a function of $\mu/ \LMS$ for $T=0$. Bottom: $M_{\rm eff}/m$ as a function of $T/ \LMS$ for $\mu=0$.}
\label{Meff}
\end{figure}

In Figure \ref {PvsgDEx} we analyze the pressure as a function of the coupling for low ($0.5 m$) and high ($5 m$) values
of $\mu$ and $T$. As expected, all methods agree with the free gas case when $g \to 0$. 
Also, at high $T,\mu$ values MFT and OPT results tend towards the free gas since one approaches the Stefan-Boltzmann limit,
while PT has a completely different behavior in this situation.
Finally, let us study the behavior of the OPT and MFT effective masses as functions of $T$ and $\mu$, as shown in Figure \ref{Meff} for $g=\pi$. 
Both quantities have a quantitatively as well as qualitatively different behavior  at small $T,\mu$ values but, as seen in the previous figures, 
this effect does not manifest itself in the pressure, probably being compensated by the presence (absence) of exchange terms within the OPT (MFT).
As $T$ or $\mu$ increases the qualitative behavior of both effective masses becomes the same although there are still quantitative differences.
These results clearly illustrate how PT is not appropriate to deal with this kind of situation since it cannot resum the condensate which arises from the interaction between the scalar field and fermions and which is related to symmetry breaking. Our application nicely illustrates the reliability of the OPT results since they exactly agree with MFT at large-$N$ (when exchange terms are neglected), while allows us to improve over this approximation by resumming exchange contributions.

\section{Results in the absence of a scalar condensate }

Let us now follow Ref. \cite {Palhares:2008yq} and neglect the one-particle-reducible two-loop diagrams containing tadpoles. In this case, which is relevant for the situation where symmetry breaking is not present, one expects that PT will perform better than in the previous case where the tadpole contributions have been considered, especially in the zero-temperature limit. Clearly MFT is not applicable in this situation, and we will restrain our analysis to OPT and PT results. At this stage, the reader should be convinced that, as explicitly shown in the previous section, the former method is able to generate nonperturbative results which can be used to access the eventual breakdown of PT. 

\subsection{Cold and dense case with $m_\phi=0$\label{mphi=0}}

In this subsection, the order-$\delta$ OPT results will be compared to the second-order perturbative predictions of Ref. \cite {Palhares:2008yq}.
The case $m_\phi=0$ is of particular interest since in this situation one can write the thermodymanic potential 
$\Omega_Y=\Omega_Y^{{\rm vac}}+\Omega_Y^{{\rm med}}(T=0,\mu)$ in a simple and compact analytic form, in terms of the vacuum part
\begin{eqnarray}
\lim_{m_{\phi}\to 0}~\Omega_Y^{{\rm vac}}
&=&
4N_F B^{{\rm REN}}(m)+\lim_{m_{\phi}\to 0}~\Omega_{{\rm vac}}^{{\rm exc,REN}}(m,m_{\phi}) \, ,
\end{eqnarray}
with
\begin{eqnarray}
\lim_{m_{\phi}\to 0}~\Omega_{{\rm vac}}^{{\rm exc,REN}}(m,m_{\phi})
&=&
N_F \frac{g^2}{2} \frac{m^4}{64\pi^{4}}
\Bigg\{ 
v_1\left( 0 \right)
+\left[ \gamma +\log\left( \frac{\Lambda^2}{m^2}\right)
 \right]v_2\left( 0 \right)
+
\frac{1}{2}\left[ \gamma +\log\left( \frac{\Lambda^2}{m^2}\right)\right]^2
v_3\left( 0 \right)
+
6~\alpha\left(m^2\right)
\Bigg\} \, ,
\end{eqnarray}
and the in-medium contribution
\begin{eqnarray}
\lim_{m_{\phi}\to 0}~\Omega_Y^{{\rm med}}(T=0,\mu)
&=&
- N_F\frac{1}{24\pi^2}
\left[
2~\mu p_f^3-3 m^2~u
\right]
-N_F\frac{g^2}{64\pi^4}
\left\{
3~u^2-4~p_f^4
+
m^2~
u
\left[ 7-3\log\left( 
\frac{m^2}{\Lambda^2} \right) \right]
\right\}
\, , \label{OmegaYResT0mphi0}
\end{eqnarray}
where $p_f^2=\mu^2-m^2$ and $u=\mu p_f-m^2 \log\left(\frac{\mu+p_f}{m}\right)$.

\begin{figure}[htb]
\includegraphics[width=8cm]{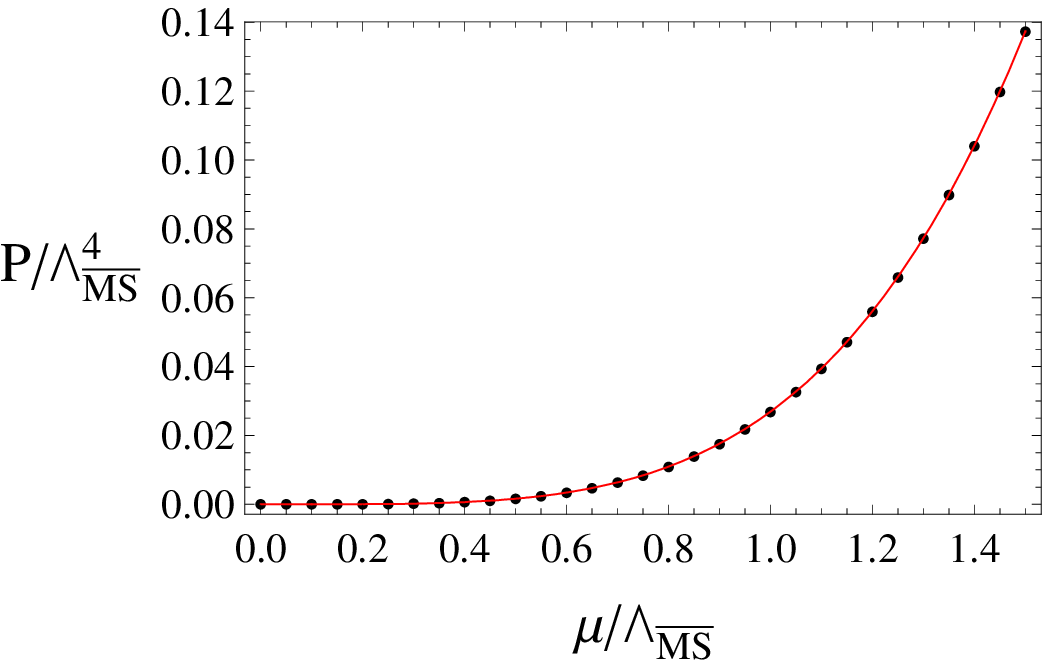}
\caption{Pressure in the absence of tadpoles normalized by $\LMS^{4}$ as a function of the chemical potential $\mu$ normalized 
by $\LMS$. Dots represent the OPT result and the line stands for PT. The fermion mass 
is fixed at $m=0.1\LMS$, and $g=\pi$.}
\label{fig2}
\end{figure}

Figure \ref{fig2} shows the pressure as a function of $\mu$ for a large value of the coupling, $g=\pi$.
As one can see both methods predict very similar results.

\vspace{0.4cm}
%
%
%

\begin{figure}[htb]
\includegraphics[width=8cm]{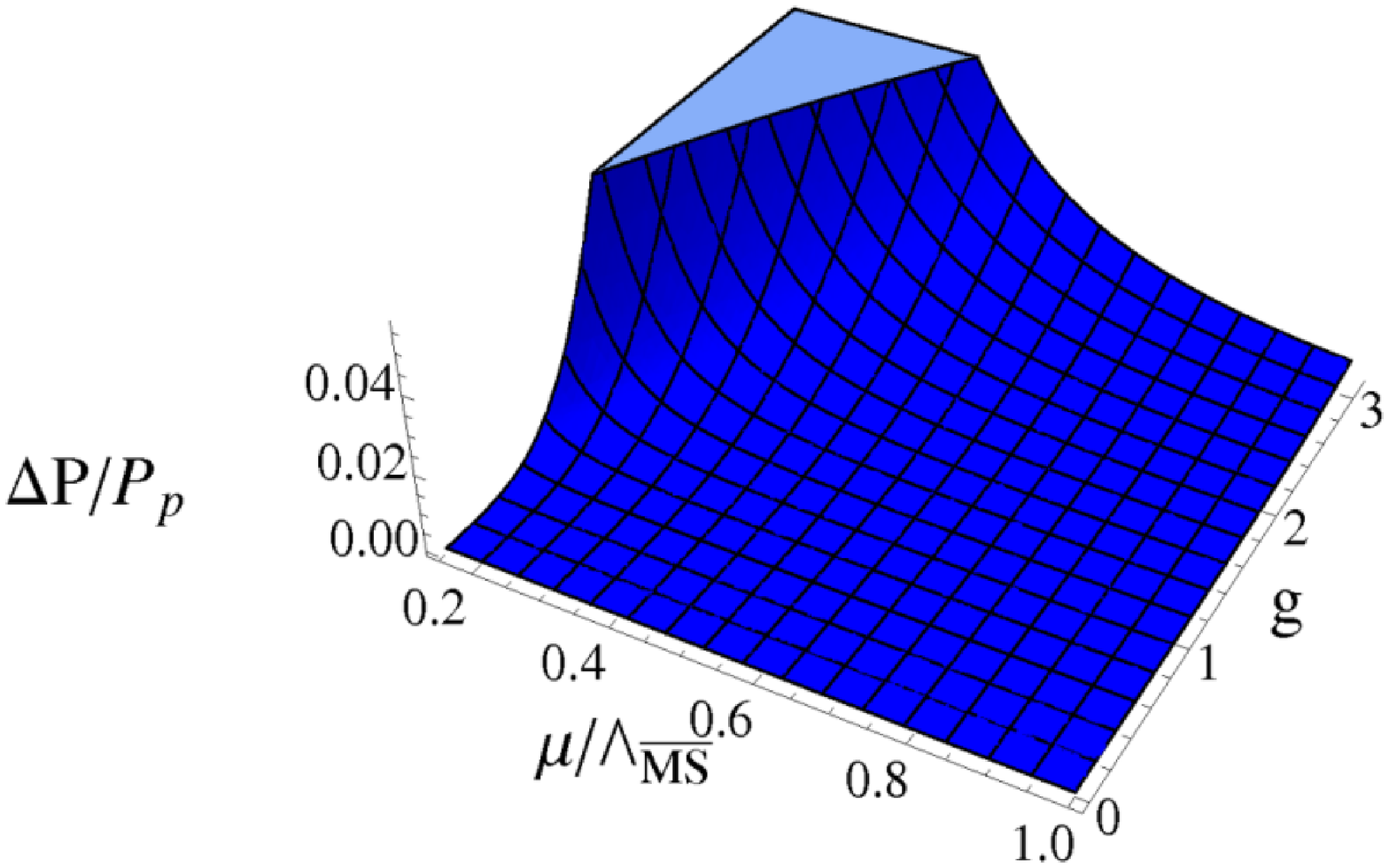}
\caption{Difference between the perturbative and the OPT pressures in the absence of tadpoles as a 
function of the coupling $g$ and the fermion chemical potential $\mu$ normalized by $\LMS$.
The fermion mass is fixed at $m=0.1\LMS$.}
\label{dif3dT0}
\end{figure}

To analyze the tiny differences, let us define the quantity $\Delta P/P_p= |(P_{\rm opt} - P_p)|/P_p$ 
where $P_p$ and $P_{\rm opt}$ are, respectively, the pressures predicted by PT and OPT.
Although the numerical discrepancies appear to be rather small, Figure \ref{dif3dT0} nicely illustrates that  $\Delta P/P_p$ increases with higher couplings and decreases with higher chemical potential values as opposed to the case where the scalar tadpole is  present. 
%
%

\vspace{0.4cm}

\hspace{2cm}

\begin{figure}
\includegraphics[width=8cm]{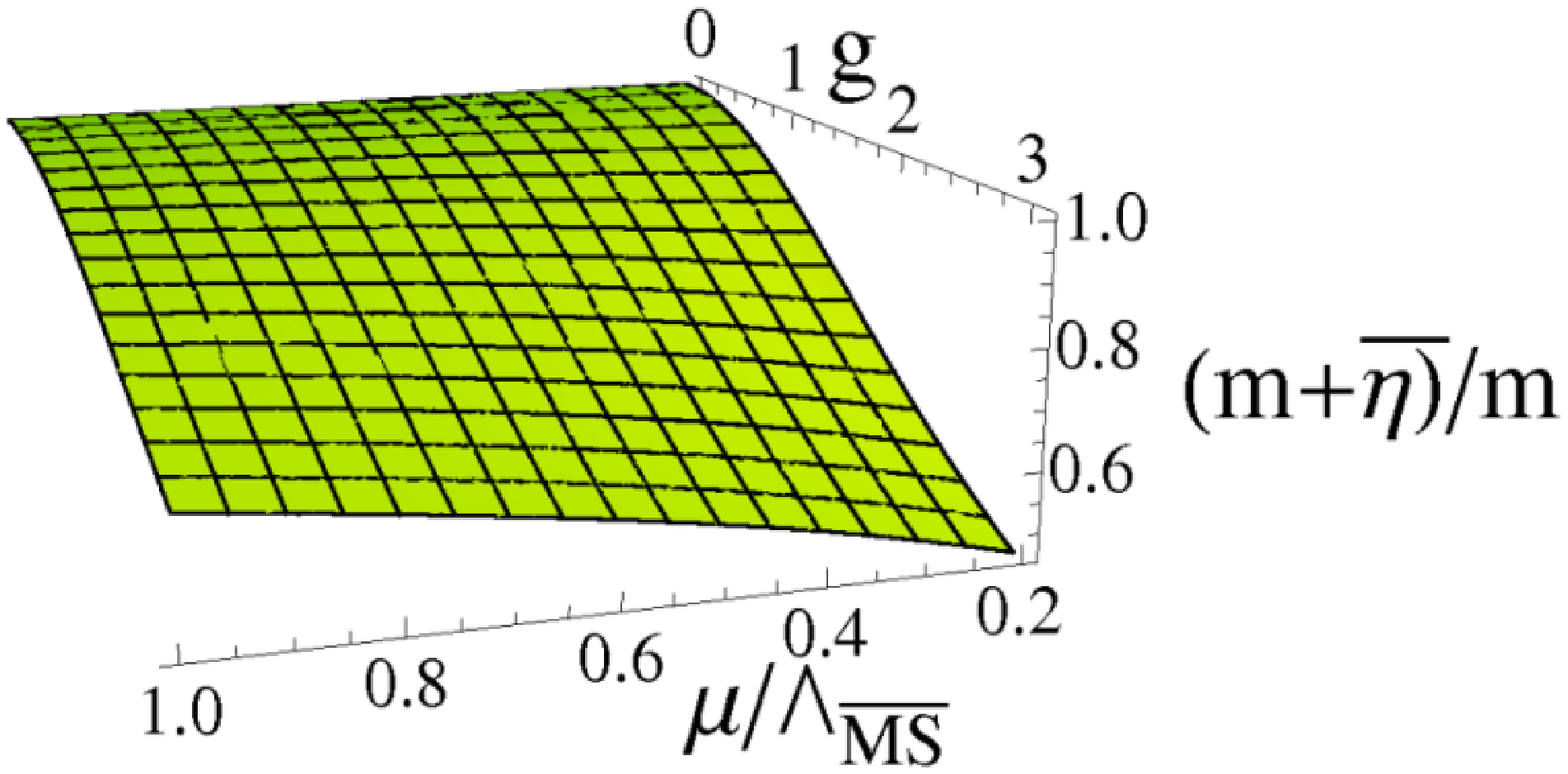}
\caption{The OPT effective mass $m+{\bar \eta}$ in units of $m$ as a 
function of the coupling $g$ and the fermion chemical potential $\mu$ normalized by $\LMS$ when tadpoles are absent.
The fermion mass is fixed at $m=0.1\LMS$.}
\label{mass3dT0}
\end{figure}

This behavior can be better understood if one analyzes how the OPT effective mass varies 
with $g$ and $\mu$. Figure \ref {mass3dT0} shows that the quantity $(m+{\bar \eta})/m$ deviates 
from $1$ as $g$ increases but, contrary to the case where tadpoles are present (see Fig. \ref {Meff}), 
approaches $1$ as $\mu$ increases.

%

Therefore, the nonperturbative OPT results support the PT results of Ref. \cite {Palhares:2008yq}, at $T=0$, when scalar condensates are not considered. Notice that even though we have considered in this subsection only the $m_\phi=0$ case our numerical simulations show that the agreement between PT and OPT remains valid for $m_\phi \ne 0$. For more details concerning the effects of the scalar mass the reader is referred to Ref. \cite {Palhares:2008yq}.

\subsection {Thermal effects}

\begin{figure}[htb]
\includegraphics[width=8cm]{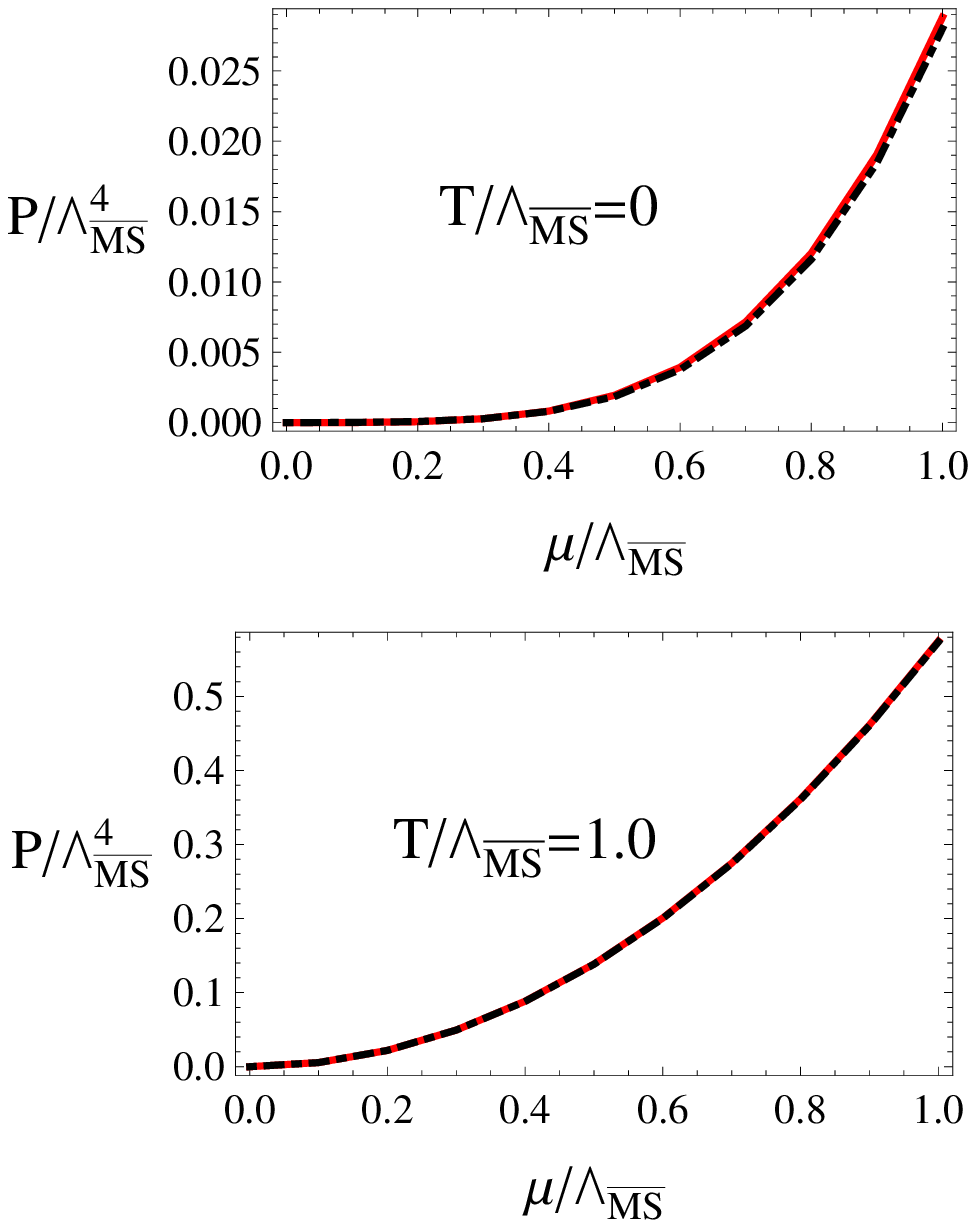}
\caption{The pressure, $P/\LMS^4$  as a function of $\mu/ \LMS$ at $T=0$ (top figure) and at $T=1.0 \LMS$  when tadpoles are absent.  PT  results are represented by the continuous lines while the OPT results are represented by the  dot-dashed lines. }
\label{SoEx}
\end{figure}

So far, we have seen through comparison with MFT and OPT that PT does not give reliable results when condensates are present but, as seen in the previous subsection, the situation improves in their absence, at least at $T=0$. In principle, it is not obvious that PT will furnish reliable results at high temperatures, even when only exchange contributions are considered. The aim of this subsection is to analyze this situation.
Figure \ref {SoEx} compares the OPT and PT results for the pressure as a function of $\mu$ in two extreme situations: $T=0$ and $T=10 m= 1.0 \LMS$. In the first case both methods agree well, as one should expect from the discussion performed in the previous section, but more surprisingly is the 
high-temperature result where OPT also supports PT.

\begin{figure}[htb]
\includegraphics[width=8cm]{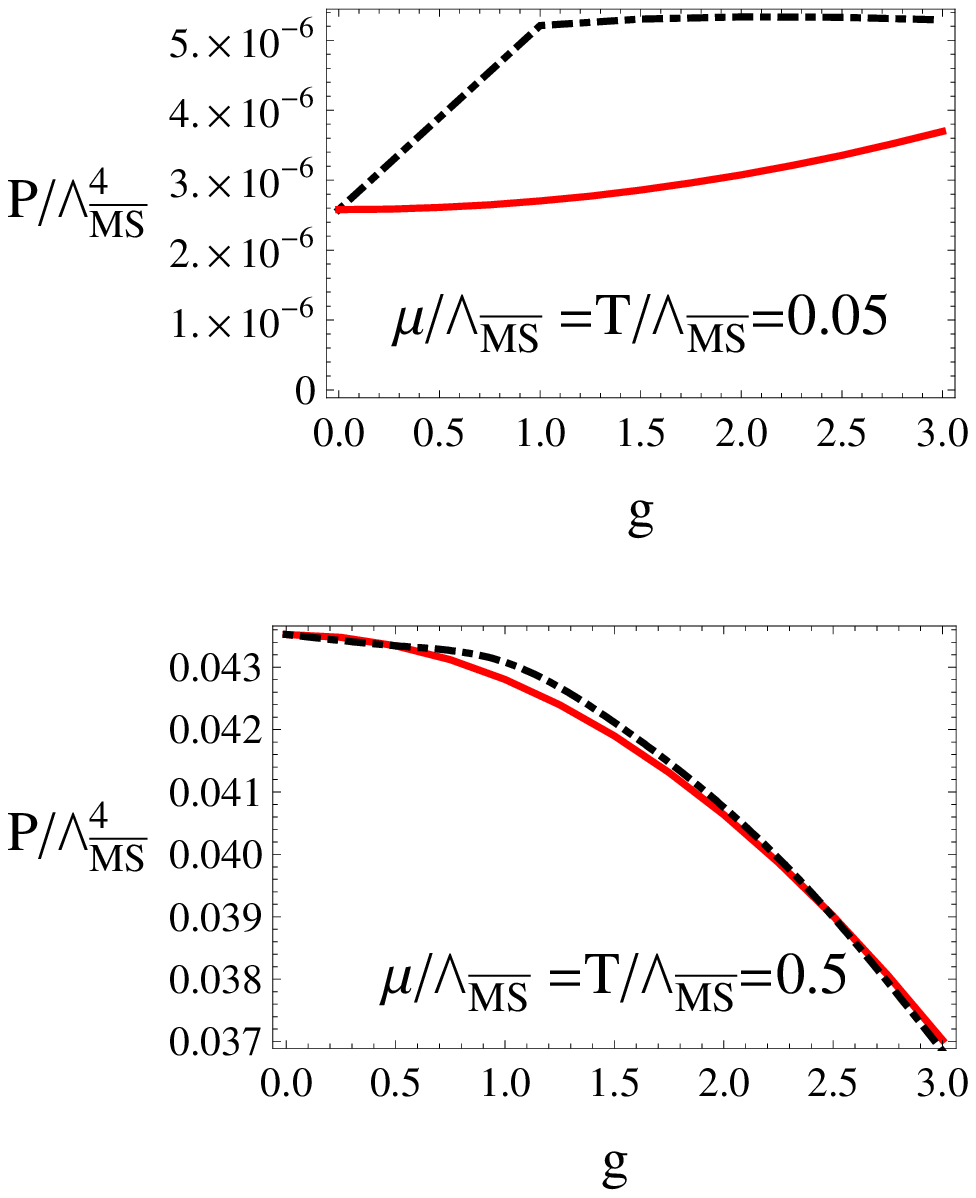}
\caption{The pressure, $P/\LMS^4$  as a function of $g$ for $\mu/=T= 0.05 \LMS$ (top figure) and at $\mu=T=0.5 \LMS$ when tadpoles are absent.  PT  results are represented by the continuous lines while the OPT results are represented by the  dot-dashed lines. }
\label{SoExg}
\end{figure}
%
%

Like in the previous $T=0$ case, the quantity $\Delta P/P_p$ assumes very small values which increase with high values of $g$ but decrease with high values of $T$ and/or $\mu$. The OPT effective mass in units of $m$, $(m+{\bar \eta})/m$, also behaves as in the previous case, deviating from $1$
as $g$ increases and, contrary to the case where tadpoles are present (see Fig. \ref {Meff}), approaching $1$ as $\mu$ and/or $T$ increase.
Therefore, it looks like the absence of direct terms means that $m \simeq m+{\bar \eta}$ at high $T$ and/or $\mu$, so that PT behaves like OPT.
Finally, Figure \ref {SoExg} suggests that even at relatively high $T$ and $\mu$ OPT and PT deviate from the free gas as the coupling increases, while in the case with condensates (see Fig. \ref {PvsgDEx} ) OPT (and MFT) had a better agreement with the free gas as opposed to PT. Therefore, the correct resummation of condensates seems to reduce the effects of interactions at high $T$ and $\mu$.

\section{Conclusions}

We have investigated the thermodynamics of the Yukawa model at finite temperature and chemical potential by evaluating its thermodynamic potential up to the two-loop level which includes direct (Hartree-like) as well as exchange (Fock-like) types of contributions. Three different methods have been considered: the usual perturbation theory (PT) with its bare mass, the optimized perturbation theory (OPT) with its effective mass given by the PMS variational criterion, and the well-known MFT with its self-consistent effective mass. 

We have considered coupling values ranging from $g=0$ to $g > > 1$ in situations where the temperature and chemical potential ranged from zero to ten times the highest mass value, which is kept fixed. As discussed in the Introduction, the Yukawa model usually emerges in the description
of various physical situations, ranging from low-energy condensed matter phenomena to 
extremely energetic QCD matter. In our approach, the characteristic features of each
physical system will be brought about essentially by the specific values of the coupling $g$ 
and the three energy scales: the fermion and
scalar masses $m$ and $m_{\phi}$, respectively, and the renormalization scale $\LMS$,
which normalizes all the quantities in our plots. As observed previously using 
the perturbative method \cite{Palhares:2008yq}, variations of the masses can significantly affect
the thermodynamic potential of the Yukawa theory, yielding
extremely different thermodynamical pictures. The renormalization scale $\LMS$ sets
the typical energy scale of the system of interest. In hadronic physics, for example,
it is reasonable to choose $\LMS\sim 1$ GeV, the confinement scale.
In this paper, we keep the discussion in general grounds, normalizing physical quantities
by $\LMS$ and fixing the masses as $m=0.1~\LMS$ and $m_{\phi}=0 \, , \, 0.5~m$, and 
concentrating on the effects of Hartree- and Fock-like interactions and nonperturbative corrections.

First, we have analyzed the pressure with both direct and exchange contributions, with the former being associated to the presence of a scalar condensate driven by the interactions with fermions. We have shown that OPT and MFT are identical if one does not consider exchange terms, in agreement with many other applications \cite {optmft}, which is reassuring since in the limit of direct contributions only the MFT resummation can be considered as ``exact''. Moreover, nonperturbative effects of exchange  contributions are readily incorporated by considering OPT consistently up to two loops. In the light of the nonperturbative approaches OPT and MFT, our results show that, as expected, naive PT is inadequate to deal with this situation since it has no ability to resum tadpoles. 

As a byproduct of this application we could see how the resummation of exchange terms performed by OPT corrects the  MFT framework, 
which corresponds to the leading order of a $1/N$ type of approximation. Having established the reliability of the OPT, we have 
followed Ref. \cite {Palhares:2008yq} imposing the absence of tadpoles at $T=0$.  We have then shown that, in this case, the OPT results turn out to be very similar to the ones given by PT. Finally, still in the limit in which condensates are not present, we have investigated the high-temperature case where one could expect the breakdown of PT. However, our results have shown that this is not the case and the numerical differences between the OPT and PT results are very small. Our results also suggest that, at high $T$ and $\mu$, the presence of condensates minimize the effects due to interactions when these contributions are properly resummed. 

One should notice that the direct application of these results as effective-model predictions 
in different physical contexts is restricted 
to definite energy regimes, within which the relevant physical degrees of freedom can be translated into
a Yukawa model. For instance, when discussing the thermodynamics of cold and dense baryonic matter,
the Yukawa model considered here might be a suitable effective theory only at small values
of the chemical potential, e.g. at $\mu\lesssim 350$ MeV. To describe the properties of baryonic matter
at higher values of $\mu$, a different framework is necessary to account for the phenomenon of color superconductivity
(for a review, see Ref. \cite{Alford:2007xm}). One alternative could be to extend this model
by adding boson fields from another, non-singlet, multiplet of the color $SU(3)$ group.
Nevertheless, since Yukawa-type interactions are almost ubiquituous in the description
of fermionic matter, the analysis in the present paper provides an understanding
of the interplay between direct and exchange contributions to the thermodynamics as well
as the role played by nonperturbative effects in the scalar Yukawa sector of any given 
extended model.

\section*{Acknowledgments} 
We thank R. L. S. Farias and R. O. Ramos for fruitful discussions. 
This work was partially supported by CAPES, CNPq, FAPERJ and FUJB/UFRJ.


\appendix


\section{Vacuum thermodynamic potential and renormalization\label{ApRen}}

In this appendix, we address the details involved in the explicit derivation
of the vacuum contributions to the two-loop thermodynamic potential of the Yukawa
theory. In particular, we concentrate on the calculation and renormalization of the 
1-loop bubble diagrams and the exchange diagram in the vacuum. The vacuum and in-medium
direct contributions (the third diagram in Fig. \ref{OmegaY-fig}), which were not considered
in Ref. \cite{Palhares:2008yq}, are left for the next appendix.

After computing the Matsubara sums (cf. Ref. \cite{Palhares:2008yq} and the next appendix), 
the pieces which are not explicitly dependent on $T$ and/or $\mu$ correspond to the vacuum 
contribution in Eq. (\ref{OmegaY(0,0)}).

The first two terms in Eq. (\ref{OmegaY(0,0)}) can be expressed in terms of the following
UV-divergent function
\beqa 
B(M) 
&=&-\frac{1}{2}\int dM~M \int\frac{d^3{\bf p}}{(2\pi)^3} 
\frac{1}{\sqrt{{\bf p}^2+M^2}}
\, .
\eeqa 
Those divergences are cancelled by a field-independent counterterm in the Lagrangian,
commonly known as a vacuum expectation value subtraction or a cosmological constant.
Within the $\overline{\textrm{MS}}$ subtraction scheme, 
the tridimensional momentum integral above is renormalized to \cite{Caldas:2000ic}:
\beqa 
B^{\textrm{REN}}(M) &=& \frac{M^4}{64\pi^2}\left[\frac{3}{2}+\log\left(  
\frac{\Lambda^2}{M^2}\right)\right]
\, .
\eeqa 

\begin{figure}[h]
\includegraphics[width=8cm]{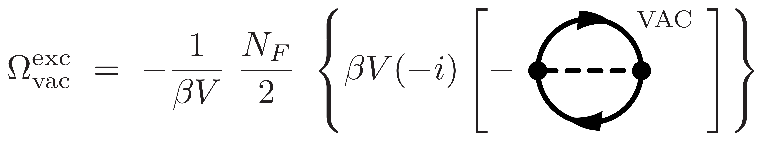}
\caption{Vacuum contribution of the exchange term written in terms of the associated vacuum diagram.}
\label{A1}
\end{figure}

The two-loop $T,\mu$-independent exchange contribution to the thermodynamic potential $\Omega_Y$, 
the last term in Eq. (\ref{OmegaY(0,0)}), can also be written in terms of UV-divergent vacuum integrals:
\beqa 
\Omega_{{\rm vac}}^{{\rm exc}}
&=&  N_F\frac{g^2}{2} 
\int\frac{d^4Pd^4Q}{(2\pi)^8}~
\frac{4(m^2+P\cdot Q)}{\left( Q^2-m^2 \right) \left( P^2-m^2 \right) \left[ (Q-P)^2-m_{\phi}^2 \right]} 
\, ,\label{Omvac2fRF}
\eeqa 
corresponding to the vacuum exchange diagram \footnote{
Throughout the appendices, whenever we refer to vacuum diagrams, we adopt the Feynman rules
from Ref. \cite{Peskin:1995ev}, with factors $(-N_F)$ associated with fermion loops excluded from the diagram definition.
}, as shown in Fig. \ref{A1}.

%

\begin{figure}[htb]
\includegraphics[width=17.5cm]{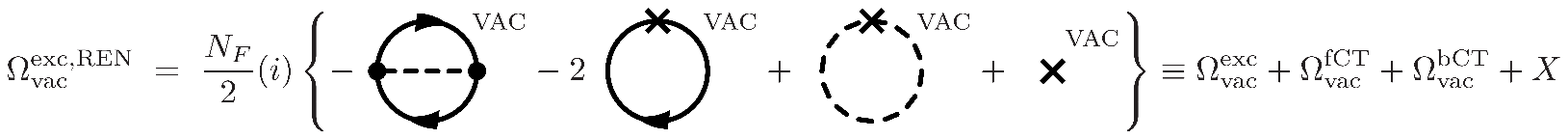}
\caption{Renormalized exchange contribution to the vacuum thermodynamic potential. The crosses indicate
counterterm vertices.}
\label{A2}
\end{figure}

The renormalization is then implemented through the usual procedure, with the addition of the
appropriate diagrams containing counterterms, as represented in Fig. \ref{A2}.
%

\begin{figure}[htb]
\includegraphics[width=7cm]{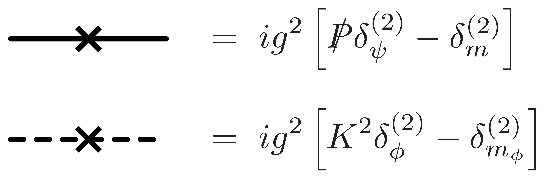}
\caption{Definition of counterterm vertices.}
\label{A3}
\end{figure}

The counterterm vertices are defined in Fig. \ref{A3} in terms of wavefunction and mass counterterms.
At this order within the $\overline{\textrm{MS}}$ subtraction scheme, these counterterm vertices
cancel exactly the 1-loop vacuum self-energy poles, yielding $(d=4-\epsilon)$:
\begin{eqnarray}
\delta_{\psi}^{(2)}=
-\frac{1}{2(4\pi)^2}~\frac{2}{\epsilon}
\quad &;& \quad
\delta_{m}^{(2)}
\frac{m}{(4\pi)^2}~\frac{2}{\epsilon}
\label{Adelpsim}
\, ,
\\
\delta_{\phi}^{(2)}=
-\frac{2}{(4\pi)^2}~\frac{2}{\epsilon}
\quad &;& \quad
\delta_{m_{\phi}}^{(2)}=
-\frac{12m^2}{(4\pi)^2}~\frac{2}{\epsilon}
\label{Adelphimphi}
\, .
\end{eqnarray}

%

%

Therefore, the second and third contributions in Fig. \ref{A2}, the vacuum bubble diagrams
with counterterm insertions, yield respectively:

\begin{eqnarray}
\Omega_{\textrm{vac}}^{\textrm{fCT}}
&=&
i\frac{N_F}{2}(-2) \int\frac{d^4P}{(2\pi)^4} ~\textrm{Tr}\left[
\left(ig^2[\slashed{P}\delta_{\psi}^{(2)}-\delta_{m}^{(2)}]
\right)
\frac{i}{\slashed{P}-m}
\right]
\nonumber\\
&=&
i\frac{N_F}{2}(-2)i(ig^2)~4\left[
m^2\delta_{\psi}^{(2)}-m\delta_{m}^{(2)}\right]\mathcal{I}(m^2)
\,,\label{AdiagfermionCTI}
\\
\Omega_{\textrm{vac}}^{\textrm{bCT}}
%
&=& 
i\frac{N_F}{2}\int\frac{d^4K}{(2\pi)^4} ~
\left(
ig^2(K^2\delta_{\phi}^{(2)}-\delta_{m_{\phi}}^{(2)})
\right)
\frac{i}{K^2-m_{\phi}^2}
\nonumber\\
&=& 
i\frac{N_F}{2}
i(ig^2)~
\left(
m_{\phi}^2\delta_{\phi}^{(2)}-\delta_{m_{\phi}}^{(2)}
\right)\mathcal{I}(m_{\phi}^2)
\, ,\label{AdiagbosonCTI}
\end{eqnarray}
where we have defined the following divergent integral
\begin{eqnarray}
\mathcal{I}(m^2)\equiv \int\frac{d^4P}{(2\pi)^4} \frac{1}{P^2-m^2}
\, ,
\end{eqnarray}
whose regularization yields
\begin{eqnarray}
\mathcal{I}^{\textrm{REG}}(m^2)
&=&
\frac{i m^2}{(4\pi)^{2}}
\left\{
\frac{2}{\epsilon}+1+
\log \left(
\frac{\Lambda^2}{m^2}
\right)
+\epsilon~
\alpha(m^2)
+O(\epsilon^2)
\right\}
\, ,\nonumber\\\label{AIreg}
\end{eqnarray}
with
\begin{eqnarray}
\alpha(m^2)\equiv \left[
\frac{1}{4}\left\{
1+
\log \left(
\frac{\Lambda^2}{m^2}
\right)
\right\}^2
+\frac{1}{4}\left(1+\frac{\pi^2}{6}\right)
\right]
\, .
\end{eqnarray}

Using the results (\ref{Adelpsim}), (\ref{Adelphimphi}) and (\ref{AIreg}) 
in (\ref{AdiagbosonCTI}) and (\ref{AdiagfermionCTI}), we arrive at the following final regularized 
expressions for the second and third terms in Fig. \ref{A2}, respectively:
\begin{eqnarray}
\Omega_{\textrm{vac}}^{\textrm{fCT}}
&=&
i\frac{N_F}{2}(-12)ig^2~\frac{m^4}{(4\pi)^{4}}
\left\{
\left(\frac{2}{\epsilon}\right)^2+\frac{2}{\epsilon}+
\frac{2}{\epsilon}\log \left(
\frac{\Lambda^2}{m^2}
\right)
+2~
\alpha(m^2)
+O(\epsilon)
\right\}
\,,\label{AdiagfermionCTF}
\\
\Omega_{\textrm{vac}}^{\textrm{bCT}}
&=& i\frac{N_F}{2}2ig^2~\frac{m_{\phi}^4}{(4\pi)^{4}}
\left[
1
-
\frac{6m^2}{m_{\phi}^2}~
\right]
\left\{
\left(\frac{2}{\epsilon}\right)^2+\frac{2}{\epsilon}+
\frac{2}{\epsilon}\log \left(
\frac{\Lambda^2}{m_{\phi}^2}
\right)
+2~
\alpha(m_{\phi}^2)
+O(\epsilon)
\right\}
\, .\label{AdiagbosonCTF}
\end{eqnarray}
%

In Fig. \ref{A2}, the contribution still to be calculated explicitly in a regularized form is
$\Omega_{\textrm{vac}}^{\textrm{exc}}$.
Using the identity
\beqa 
m^2+P\cdot Q &=& 
2m^2-\frac{1}{2}m_{\phi}^2 -\frac{1}{2}\left[ (Q-P)^2-m_{\phi}^2 \right]
+\frac{1}{2}(P^2-m^2)+\frac{1}{2}(Q^2-m^2) \, ,
\eeqa 
we can rewrite $\Omega_{{\rm vac}}^{{\rm exc}}$, Eq. (\ref{Omvac2fRF}), as:
\beqa 
\Omega_{{\rm vac}}^{{\rm exc}} =  N_F 2g^2 \left\{ \left( 2m^2-\frac{1}{2}m_{\phi}^2  \right)
~I_1^{b} -\frac{1}{2} ~I_2^{b} +I_3^{b}\right\}
\, , \label{Omvac2fIs}
\eeqa 
in terms of the following integrals:
\beqa 
I_1^{b} &\equiv& \int\frac{d^4Pd^4Q}{(2\pi)^8}~
\frac{1}{\left( Q^2-m^2 \right) \left( P^2-m^2 \right) \left[ (Q-P)^2-m_{\phi}^2 \right]} 
\, ,
\label{i1b} \\ 
I_2^{b} &\equiv& \int\frac{d^4Pd^4Q}{(2\pi)^8}~
\frac{1}{\left( Q^2-m^2 \right) \left( P^2-m^2 \right) } 
\, ,
\label{i2b} \\
I_3^{b} &\equiv& \int\frac{d^4Pd^4Q}{(2\pi)^8}~\frac{1}{\left( Q^2-m^2 \right)  \left[ (Q-P)^2-m_{\phi}^2 \right]}
\, .
\label{i3b}
\eeqa 
Defining
\beqa 
J(a,b) &\equiv& \int \frac{d^4Pd^4Q}{(2\pi)^8} \frac{1}{(P^2-a)(Q^2-b)} \, ,
\eeqa 
we have $I_2^b=J(m^2,m^2)$ and $I_3^b=J(m^2,m_{\phi}^2)$. The dimensional
regularization of $J(a,b)$ is straightforward, yielding:
\beqa 
J^{\textrm{REG}}(a,b) &=& 
-\frac{1}{(4\pi)^d}~\left( \frac{e^{\gamma}\Lambda^2}{4\pi} 
\right)^{\epsilon} ~
\left[\Gamma\left( 1-\frac{d}{2} \right) \right]^2
~(ab)^{\frac{d}{2}-1} \, ,\label{JREGres}
\eeqa 
where $d=4-\epsilon$.

On the other hand, the evaluation of the integral $I_1^b$ is extremely involved,
essentially due to the absence of factorization of terms containing different masses.
This calculation was performed in Ref. \cite{Davydychev:1992mt} and the result is
\beqa 
I_1^{b,\textrm{REG}} &=& \left( \frac{e^{\gamma}\Lambda^2}{4\pi} \right)^{\epsilon} ~
\frac{\pi^{4-\epsilon}}{(2\pi)^{2d}} 
\left(m^2\right)^{1-\epsilon}
A\left(\frac{\epsilon}{2}\right)
\Big\{ -\frac{4}{\epsilon^2}(1+2z)+\frac{2}{\epsilon}\left[ 4z~\log(4z) \right]-
\nonumber \\
&&-
2z~\left[ \log(4z) \right]^2+
2(1-z)\Phi(z) +O(\epsilon)
\Big\}  \, , \label{i1bREGres}
\eeqa 
with $z \equiv \frac{m_{\phi}^2}{4m^2}$,
\begin{eqnarray}
A\left(\frac{\epsilon}{2}\right)&\equiv& 
\frac{\left[ \Gamma(1+\epsilon/2) \right]^2}{(1-\epsilon/2)(1-\epsilon)}
= 1+\epsilon~\beta_1+\epsilon^2\beta_2+O(\epsilon^3) \label{A}\\
\beta_1 &\equiv& \frac{3}{2}-\gamma \\
\beta_2 &\equiv& \frac{7}{4}-\frac{3}{2}\gamma+\frac{1}{2}\gamma^2+\frac{\pi^2}{24}
\, ,
\\
\Phi(z) &\equiv& 4z~\Bigg\{ \left[ 2-\log(4z) \right]~{}_2F_1\left(1,1,\frac{3}{2};~z\right)
-\left.\left[ \frac{\partial}{\partial a}~{}_2F_1\left(a,1,\frac{3}{2};~z\right)
\right] \right|_{a=1}
-\nonumber\\ &&
-\left.\left[ \frac{\partial}{\partial c}~{}_2F_1 \left( 1,1,c;~z \right)
\right] \right|_{c=\frac{3}{2}} \Bigg\}
\, ,
\end{eqnarray}
and ${}_2F_1$ is the hypergeometric function, defined by:
\begin{equation}
{}_2F_1(a,b,c; ~z) \equiv \sum_{k=0}^{\infty}\frac{(a)_k(b)_k}{(c)_k}~\frac{z^k}{k!} 
\, ,
\end{equation}
where $(a)_k \equiv \frac{\Gamma(a+k)}{\Gamma(a)}$ is the Pochhammer symbol. 
In Ref. \cite{Davydychev:1992mt}, simplified expressions for $\Phi(z)$ valid
in the regions $z>1$ or $z\le 1$ were also derived.

Taking the results in Eq. (\ref{JREGres}) and in Eq. (\ref{i1bREGres}) into the
Eq. (\ref{Omvac2fIs}) and expanding around $\epsilon=0$, one obtains, after a long algebra:
\begin{eqnarray}
\Omega_{\textrm{vac}}^{{\rm exc}}
&=&  
N_F \frac{g^2}{2} \frac{m^4}{64\pi^{4}}~
\Bigg\{
\Big[ \textrm{poles in}~ \epsilon=0 \Big]+ 
v_1\left( \frac{m_{\phi}^2}{4m^2} \right)
+\left[ \gamma +\log\left( \frac{\Lambda^2}{m^2}\right)
 \right]~v_2\left( \frac{m_{\phi}^2}{4m^2} \right)
+\nonumber\\
&&
\quad+ \frac{1}{2}~\left[ \gamma +\log\left( \frac{\Lambda^2}{m^2}\right)\right]^2
~v_3\left( \frac{m_{\phi}^2}{4m^2} \right)
+O(\epsilon)
\Bigg\}
\, , \label{apOmvac2fResFinal}
\end{eqnarray}
where
\begin{eqnarray}
v_1(z) &=& 2(\gamma_0-4\beta_2)-8z~(2\gamma_0+\beta_2)+16\beta_2~z^2 +4(1-z)^2~\Phi(z)
+\nonumber \\
&&\quad
+ \log(4z)~\left\{ 8z~\left[ 2( 1-z )~\beta_1 + \gamma_{1} \right] \right\}+
\left[ \log(4z) \right]^2~\left\{ 4z^2-6z \right\} \, ,\label{apv1}
\\ 
v_2(z)&=& 2(\gamma_1-4\beta_1)-8z(2\gamma_1+\beta_1)+16\beta_1~z^2+
8z(3-2z)~\log(4z) \, , \\ 
v_3(z) &=& -6-24~z+16~z^2 \, ,\label{apv3}
\end{eqnarray}
with
\beqa 
\gamma_0 &\equiv& \frac{3}{4}+\frac{\gamma(\gamma-2)}{2}+\frac{\pi^2}{24}
\, ,
\\
\gamma_{1} &\equiv& 1-\gamma
\, .
\eeqa 

Finally, the 2-loop contribution to the vacuum expectation value counterterm, $X$,
is then defined in order to cancel the poles in Eqs. (\ref{AdiagfermionCTF}),
(\ref{AdiagbosonCTF}) and (\ref{apOmvac2fResFinal}).

Therefore, collecting the finite parts of Eqs. (\ref{AdiagfermionCTF}), 
(\ref{AdiagbosonCTF}) and (\ref{apOmvac2fResFinal}), we obtain
the final expression for the renormalized exchange vacuum contribution
to the thermodynamic potential:
\begin{eqnarray}
\Omega_{\textrm{vac}}^{\textrm{exc,REN}}
&=&
\frac{N_F}{2}(i)
\Bigg\{
- i g^2 \frac{m^4}{64\pi^{4}} 
\Bigg( 
v_1(z)
+\left[ \gamma +\log\left( \frac{\Lambda^2}{m^2}\right)
 \right]~
v_2(z)
+ \frac{1}{2}~\left[ \gamma +\log\left( \frac{\Lambda^2}{m^2}\right)\right]^2~
v_3(z)
\Bigg)
-\nonumber\\
&&\quad
-12ig^2~\frac{m^4}{(4\pi)^{4}}
\left[
2~
\alpha(m^2)
\right]
+ 2ig^2~\frac{m_{\phi}^4}{(4\pi)^{4}}
\left[
1
-
\frac{6m^2}{m_{\phi}^2}~
\right]
\left[
2~
\alpha(m_{\phi}^2)
\right]
\Bigg\}
\, .\label{OmegaRenF}
\end{eqnarray}
%

%

%

\section{Vacuum and in-medium direct contribution to the thermodynamic potential\label{Apdir}}

\begin{figure}[htb]
\includegraphics[width=16cm]{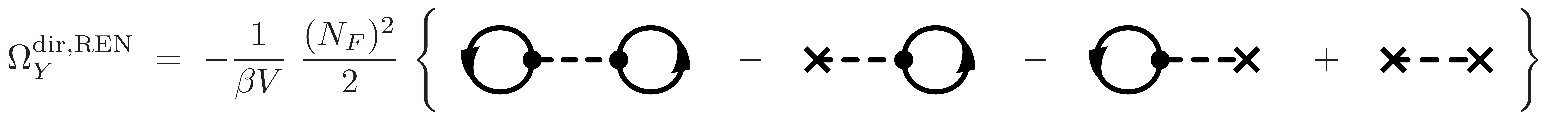}
\caption{Renormalized direct contribution to the thermodynamic potential, represented by
in-medium diagrams. As before, the crosses indicate (vacuum) counterterm vertices.}
\label{B1}
\end{figure}

The third term in the diagrammatic expansion of the perturbative thermodynamic potential,
shown in Fig \ref{OmegaY-fig}, is the direct term. In this appendix, we concentrate on the explicit evaluation
of both vacuum and in-medium parts of this contribution. As usual,
the full renormalized form of the direct term of the thermodynamic potential is obtained 
through the addition of the appropriate counterterms, as shown in Fig. \ref{B1}.

%
%
%
%
%
\begin{figure}[htb]
\includegraphics[width=6cm]{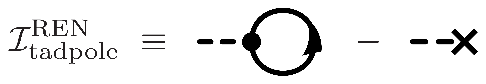}
\caption{Definition of the renormalized tadpole integral.}
\label{B2}
\end{figure}

Defining the renormalized tadpole integral, as in Fig. \ref{B2}, we can rewrite the renormalized
direct contribution to the thermodynamic potential as:
\begin{eqnarray}
\Omega^{\textrm{dir,REN}}_Y 
&=&
-\frac{1}{\beta V}~\frac{(N_F)^2}{2}~
\left\{\beta V \left[\frac{1}{m_{\phi}^2-K^2}\right]_{K^2=0} \left(\mathcal{I}_{\textrm{tadpole}}^{\textrm{REN}}\right)^2\right\}
=
-~\frac{(N_F)^2}{2m_{\phi}^2}~
\left(\mathcal{I}_{\textrm{tadpole}}^{\textrm{REN}}\right)^2
\label{Omegadir}
\, ,
\end{eqnarray}

Following the $\overline{\textrm{MS}}$ prescription, the counterterm is defined through the renormalization
of the 1-loop fermionic self-energy, cancelling exactly the pole in $\epsilon=0$ of the tadpole integral, yielding:

\begin{eqnarray}
\mathcal{I}_{\textrm{tadpole}}^{\textrm{REN}}
&=&
\Bigg\{
(-g)~\sumint_P ~\textrm{Tr}\left[\frac{1}{\slashed{P}-m}\right]
\Bigg\}^{\textrm{REN}}
\nonumber\\
&=&
\Bigg\{
(-g)~4m\int\frac{d^3{\bf p}}{(2\pi)^3}\frac{1}{2E_{{\bf p}}}(-1)
\Bigg\}^{\textrm{REN}}
+
\nonumber\\
&&+
(-g)~4m\int\frac{d^3{\bf p}}{(2\pi)^3}\frac{1}{2E_{{\bf p}}}
\left[\frac{1}{\exp[E_{{\bf p}}/T-\mu/T]+1}+ \frac{1}{\exp[E_{{\bf p}}/T+\mu/T]+1} \right]
\nonumber\\
&\equiv&
I_{\textrm{tadpole}}^{\textrm{vac,REN}}+I_{\textrm{tadpole}}^{\textrm{med}}
\, ,
\end{eqnarray}
with $E_{{\bf p}}^2={\bf p}^2+m^2$,
\begin{eqnarray}
I_{\textrm{tadpole}}^{\textrm{med}}&=&-g~2m\int\frac{d^3{\bf p}}{(2\pi)^3}\frac{1}{E_{{\bf p}}}
\left[\frac{1}{\exp[E_{{\bf p}}/T-\mu/T]+1}+ \frac{1}{\exp[E_{{\bf p}}/T+\mu/T]+1} \right]
\, ,
\end{eqnarray}
and
\begin{eqnarray}
I_{\textrm{tadpole}}^{\textrm{vac,REN}}
&=&
\Bigg\{
g~2m\int\frac{d^3{\bf p}}{(2\pi)^3}\frac{1}{E_{{\bf p}}}
\Bigg\}^{\textrm{REN}}
=
-4g~\frac{dB^{\textrm{REN}}(m)}{dm}
=-g~\frac{m^3}{(2\pi)^2}\left[
1+\log\left(
\frac{\Lambda^2}{m^2}
\right)\right]
\, .
\end{eqnarray}

Finally, taking these results into Eq. (\ref{Omegadir}),
we obtain the following expression for the renormalized direct term of the 
thermodynamic potential:
\begin{eqnarray}
\Omega^{\textrm{dir,REN}}_Y &=&
-~\frac{(N_F)^2}{2m_{\phi}^2}~
\left(\mathcal{I}_{\textrm{tadpole}}^{\textrm{REN}}\right)^2
~~=~~
-~\frac{(N_F)^2}{2m_{\phi}^2}~
\left(I_{\textrm{tadpole}}^{\textrm{vac,REN}}+I_{\textrm{tadpole}}^{\textrm{med}}\right)^2
\equiv
\Omega^{\textrm{dir,REN}}_{\textrm{vac}}+ \Omega^{\textrm{dir}}_{\textrm{med}}
\label{OmegadirRES}
\, ,
\end{eqnarray}
where
\begin{eqnarray}
\Omega^{\textrm{dir,REN}}_{\textrm{vac}}
&=&
-~\frac{(N_F)^2}{2m_{\phi}^2}~
\left(I_{\textrm{tadpole}}^{\textrm{vac,REN}}\right)^2
\\
&=&
-g^2~\frac{(N_F)^2}{2m_{\phi}^2}~
\left\{
\frac{m^3}{(2\pi)^2}\left[
1+\log\left(
\frac{\Lambda^2}{m^2}
\right)\right]\right\}^2
\, ,
\\
\nonumber
\\
\Omega^{\textrm{dir}}_{\textrm{med}}&=&
-~\frac{(N_F)^2}{2m_{\phi}^2}~
\left[2~I_{\textrm{tadpole}}^{\textrm{vac,REN}}~I_{\textrm{tadpole}}^{\textrm{med}}+
\left(I_{\textrm{tadpole}}^{\textrm{med}}\right)^2\right]
\\
&=&
-g^2T^2~\frac{(N_F)^2m^4}{(4\pi^4)m_{\phi}^2}~
\left[
1+\log\left(
\frac{\Lambda^2}{m^2}
\right)\right]
~\int~z^2~dz~\frac{1}{E_z}
\left[\frac{1}{\exp[E_{z}-\mu/T]+1}+ \frac{1}{\exp[E_{z}+\mu/T]+1} \right]
-
\nonumber\\
&&
-g^2T^4~\frac{(N_F)^2m^2}{(2\pi^4)m_{\phi}^2}~
\left\{\int~z^2~dz~\frac{1}{E_z}
\left[\frac{1}{\exp[E_z-\mu/T]+1}+ \frac{1}{\exp[E_z+\mu/T]+1} \right]
\right\}^2
\label{OmegadirRES}
\, ,
\end{eqnarray}
in terms of the dimensionless quantities $E_z^2\equiv z^2+m^2/T^2$ and $z=p/T$.



\begin{thebibliography}{99}

\bibitem{hubbard}
  T.~Baier, E.~Bick and C.~Wetterich,
  Phys. Rev. B {\bf 62}, 15471 (2000); 
  {\bf 70}, 125111 (2004); 
  Phys.\ Lett.\  B {\bf 605}, 144 (2005).
  
\bibitem{Rischke:2003mt}
 D.~H.~Rischke,
 Prog.\ Part.\ Nucl.\ Phys.\  {\bf 52}, 197 (2004).
  
\bibitem{QM}
{\it Procs. of Quark Matter 2006}, 
J. Phys. G {\bf 34} S173 (2007).  

\bibitem{stars}
N.~K. Glendenning, {\em Compact Stars --- Nuclear Physics, Particle Physics,
and General Relativity} (Springer, New York, 2000).

\bibitem{Stephanov:2007fk}
  M.~A.~Stephanov,
  PoS {\bf LAT2006}, 024 (2006).

\bibitem{Palhares:2008yq}
  L.~F.~Palhares and E.~S.~Fraga,
  Phys.\ Rev.\  D {\bf 78}, 025013 (2008).

\bibitem{Farias:2008fs}
  R.~L.~S.~Farias, G.~Krein and R.~O.~Ramos,
  Phys.\ Rev.\  D {\bf 78}, 065046 (2008).

\bibitem{GellMann:1960np}
  M.~Gell-Mann and M.~Levy,
  Nuovo Cim.\  {\bf 16}, 705 (1960).

\bibitem{Bilic:1997sh}
  N.~Bilic and H.~Nikolic,
  Eur.\ Phys.\ J.\  C {\bf 6}, 513 (1999).

\bibitem{quarks-chiral} 
L.~P.~Csernai and I.~N.~Mishustin, Phys.\ Rev.\
Lett.\ \textbf{74}, 5005 (1995); 
A.~Abada and J.~Aichelin, 
Phys.\ Rev.\ Lett.\ \textbf{74}, 3130 (1995); 
A.~Abada and M.~C.~Birse, Phys.\ Rev.\ D \textbf{55}, 6887 (1997).

\bibitem{ove} 
I.~N.~Mishustin and O.~Scavenius, 
Phys.\ Rev.\ Lett.\ \textbf{83}, 3134 (1999). 

\bibitem{Scavenius:1999zc} 
O.~Scavenius and A.~Dumitru, 
Phys.\ Rev.\ Lett.\ \textbf{83}, 4697 (1999). 

\bibitem{Caldas:2000ic}
  H.~C.~G.~Caldas, A.~L.~Mota and M.~C.~Nemes,
  Phys.\ Rev.\  D {\bf 63}, 056011 (2001).

\bibitem{Scavenius:2000qd} 
O.~Scavenius, A.~Mocsy, I.~N.~Mishustin and
D.~H.~Rischke, 
Phys.\ Rev.\ C \textbf{64}, 045202 (2001). 

\bibitem{Scavenius:2001bb} 
O.~Scavenius, A.~Dumitru, E.~S.~Fraga,
J.~T.~Lenaghan and A.~D.~Jackson, 
Phys.\ Rev.\ D \textbf{63}, 116003 (2001). 

\bibitem{Weinberg:1978kz}
 S.~Weinberg,
 Physica A {\bf 96}, 327 (1979).

\bibitem{vanKolck:1999mw}
  U.~van Kolck,
  Prog.\ Part.\ Nucl.\ Phys.\  {\bf 43}, 337 (1999).
  
\bibitem{Fraga:2001id}
E.~S.~Fraga, R.~D.~Pisarski and J.~Schaffner-Bielich,
Phys.\ Rev.\ D {\bf 63}, 121702 (2001); 
Nucl.\ Phys.\ A {\bf 702}, 217 (2002).

\bibitem{tony}
J.~P.~Blaizot, E.~Iancu and A.~Rebhan,
Phys.\ Rev.\ D {\bf 63}, 065003 (2001).

\bibitem{andersen}
J.~O.~Andersen and M.~Strickland,
Phys.\ Rev.\ D {\bf 66}, 105001 (2002).

\bibitem{Braaten:2002wi}
  E.~Braaten,
  Nucl.\ Phys.\  A {\bf 702}, 13 (2002).
  
\bibitem{Palhares:2007zz}
  L.~F.~Palhares and E.~S.~Fraga,
  Braz.\ J.\ Phys.\  {\bf 37}, 26 (2007); 
%
  Int.\ J.\ Mod.\ Phys.\  E {\bf 16}, 2806 (2007); 
%
  E.~S.~Fraga and L.~F.~Palhares,
  AIP Conf.\ Proc.\  {\bf 892}, 479 (2007).
  
\bibitem{kapusta-gale}
J. I. Kapusta and C. Gale, 
{\it Finite-Temperature Field Theory: Principles and Applications}
(Cambridge University Press, 2006).

\bibitem{OPT}
A. Okopinska, Phys. Rev. D {\bf 35}, 1835 (1987); 
M. Moshe and A. Duncan, Phys. Lett. B {\bf 215}, 352 (1988).

\bibitem{OPT2}
R. Seznec and J. Zinn-Justin, J. Math.  Phys. {\bf 20}, 1398 (1979); 
%
J. C. Le Guillou and J. Zinn-Justin, Ann. Phys. {\bf 147}, 57 (1983); 
%
V. I. Yukalov, Moscow Univ. Phys. Bull. {\bf 31}, 10 (1976); 
%
W. E. Caswell, Ann. Phys.  (N.Y) {\bf 123}, 153 (1979); 
%
I. G.  Halliday and P. Suranyi, Phys. Lett. B {\bf 85}, 421 (1979); 
%
J. Killinbeck, J. Phys. A {\bf 14}, 1005 (1981); 
%
R. P.  Feynman and H. Kleinert, Phys. Rev. A {\bf 34}, 5080 (1986); 
%
H. F. Jones and M. Moshe, Phys. Lett. B {\bf 234}, 492 (1990); 
%
A. Neveu, Nucl. Phys. (Proc.  Suppl.) B {\bf 18}, 242 (1991); 
%
V. Yukalov, J. Math. Phys {\bf 32}, 1235 (1991); 
%
C.~M. Bender et al., Phys. Rev. D {\bf 45}, 1248 (1992); 
%
S. Gandhi and M. B. Pinto, Phys. Rev. D {\bf 46}, 2570 (1992); 
%
H.  Yamada, Z. Phys. C {\bf 59}, 67 (1993); 
%
K. G. Klimenko, Z. Phys. C {\bf 60}, 677 (1993); 
%
A.N.  Sissakian, I. L. Solovtsov and O. P. Solovtsova, Phys.  Lett. B {\bf 321}, 381 (1994); 
%
H. Kleinert, Phys. Rev. D {\bf 57}, 2264 (1998); 
Phys. Lett. B {\bf 434}, 74 (1998); 
for a review, see H. Kleinert and V.  Schulte-Frohlinde, 
{\em Critical Properties of $\phi^4$-Theories}, Chap. 19 (World Scientific, Singapure 2001); 
%
K. G. Klimenko, Z. Phys. C {\bf 50}, 477 (1991)

  J.~L.~Kneur, M.~B.~Pinto and R.~O.~Ramos,
  Phys.\ Rev.\  D {\bf 74}, 125020 (2006).
%
\bibitem{Kneur:2007vj}
  J.~L.~Kneur, M.~B.~Pinto, R.~O.~Ramos and E.~Staudt,
  Phys.\ Lett.\  B {\bf 657}, 136 (2007); 
  Phys.\ Rev.\  D {\bf 76}, 045020 (2007).

\bibitem{BEC}
F. F. Souza Cruz, M. B. Pinto and R. O. Ramos, 
Phys. Rev. B {\bf 64}, 014515 (2001); 
%
E. Braaten and E. Radescu, Phys. Rev. Lett. {\bf 89}, 271602 (2002), 
Phys. Rev. A {\bf 66}, 063601 (2002); 
%
J.-L. Kneur, M. B. Pinto and R. O. Ramos, Phys. Rev. Lett.  {\bf 89}, 210403 (2002), 
Phys. Rev. A {\bf 68}, 043615 (2003); 
%
J.-L. Kneur, A. Neveu and M. B. Pinto, Phys. Rev. A {\bf 69}, 053624 (2004); 
%
J.-L. Kneur and M. B. Pinto, Phys. Rev. A {\bf 71}, 033613 (2005); 
%
B. Kastening, Phys. Rev. A {\bf 70}, 043621 (2004).

\bibitem{OPT3}
  G.~A.~Hajj and P.~M.~Stevenson,
  Phys.\ Rev.\  D {\bf 37}, 413 (1988);
%
  A.~Okopinska,
  Physica A {\bf 158}, 64 (1989);
%
  K.~G.~Klimenko,
  Z.\ Phys.\  C {\bf 43}, 581 (1989).


\bibitem{future}
E. S. Fraga, L. F. Palhares and M. Benghi Pinto, 
work in progress.

\bibitem{PRDMR}
 M.~B.~Pinto and R.~O.~Ramos, 
 Phys. Rev.  D {\bf 60}, 105005 (1999).

\bibitem{PMS}
P. M. Stevenson, Phys. Rev. D {\bf 23}, 2916 (1981); 
Nucl. Phys. B {\bf 203}, 472 (1982).

\bibitem{optmft} 
%
  S.~K.~Gandhi, H.~F.~Jones and M.~B.~Pinto,
  Nucl.\ Phys.\  B {\bf 359}, 429 (1991);
  G.~Krein, D.~P.~Menezes and M.~B.~Pinto,
  Phys.\ Lett.\  B {\bf 370}, 5 (1996).

\bibitem{Alford:2007xm}
  M.~G.~Alford, A.~Schmitt, K.~Rajagopal and T.~Schafer,
  Rev.\ Mod.\ Phys.\  {\bf 80}, 1455 (2008).

\bibitem{Peskin:1995ev}
  M.~E.~Peskin and D.~V.~Schroeder,
  {\it An Introduction To Quantum Field Theory,}
(Addison-Wesley, 1995).

\bibitem{Davydychev:1992mt}
  A.~I.~Davydychev and J.~B.~Tausk,
  Nucl.\ Phys.\  B {\bf 397}, 123 (1993).

\end{thebibliography}
\end{document}